\documentclass[conference]{IEEEtran}

\usepackage{paralist} 
\usepackage{amsmath,amsthm}
\usepackage{thm-restate}
\usepackage{amssymb}
\usepackage{epsfig} 
\usepackage{epstopdf}
\usepackage{float}
\usepackage{graphicx,algorithmic}
\usepackage{caption, subcaption}
\usepackage{wrapfig}
\usepackage{multirow}
\usepackage{multicol}
\usepackage{todonotes}
\usepackage{tikz}
\usetikzlibrary{fit} 
\usetikzlibrary{backgrounds} 
\usetikzlibrary{patterns,calc}
\usetikzlibrary{decorations.pathreplacing}
\usetikzlibrary{positioning,fit, backgrounds, calc}
\usetikzlibrary{decorations.pathmorphing}
\usetikzlibrary{decorations.markings}
\tikzstyle{background}=[rectangle,fill=gray!10, inner sep=0.1cm, rounded corners=0mm]
\usepgflibrary{shapes}
\usetikzlibrary{automata,arrows}
\usetikzlibrary{shadows}
\newcommand{\notiff}{%
  \mathrel{{\ooalign{\hidewidth$\not\phantom{"}$\hidewidth\cr$\iff$}}}}

\newtheorem{theorem}{Theorem}

\newtheorem{lemma}[theorem]{Lemma}

\newtheorem{definition}{Definition}
\newtheorem{example}{Example}

\usepackage{tikz}
\usetikzlibrary{shapes,arrows,fit,calc,positioning,automata}

\newcommand{\nonl}{\renewcommand{\nl}{\let\nl\oldnl}}

\makeatletter
\newif\if@restonecol
\makeatother

\usepackage[ruled,vlined,linesnumbered,lined]{algorithm2e}

\usepackage{xcolor,colortbl}
\definecolor{Gray}{gray}{0.95}
\definecolor{LightCyan}{rgb}{0.88,1,1}

 \usepackage{relsize}

\newcommand{\refine}[1]{\ensuremath{\lfloor {#1} \rfloor}}
\newcommand{\MCC}{\ensuremath{\mathsf{MCC}}}

\newcommand{\eat}[1]{}
\newcommand{\op}{\ensuremath{\mathsf{op}}}

\newcommand{\Ff}{F}

\renewcommand{\implies}{\Rightarrow}

\renewcommand{\iff}{\Leftrightarrow}

\newcommand{\mcomment}[1]{}

\newcommand{\Ckt}{\ensuremath{\mathcal{D}}}
\newcommand{\VSIDS}{\ensuremath{\mathsf{VSIDS}}}
\newcommand{\OutVarToBranch}{\textsc{ChooseOutputVar}}
\newcommand{\GetCkt}{\textsc{GetCkt}}
\newcommand{\GetCktWithDefs}{\textsc{GetDefCkt}}
\newcommand{\FindFDOuts}{\textsc{FindFD}}
\newcommand{\FindFDOutsAndRefine}{\textsc{FDRefine}}

\newcommand{\obfss}{\textsc{bfss}}
\newcommand{\bfss}{\textsc{bfss}}
\newcommand{\bddbfss}{\textsc{BDD}^{\textsc{bfss}}}

\newcommand{\cadet}{\textsc{Cadet}}

\newcommand{\qbf}{\textsc{QBFEval}}
\newcommand{\fdefs}{f-defs}

\newcommand{\factqd}{\textsc{FA.QD}}

\newcommand{\fact}{\textsc{Factorization}}
\newcommand{\dsharp}{\textsc{DSharp}}

\newcommand{\cnfform}[1]{\ensuremath{\varphi_{{#1}}}}

\newcommand{\bX}{\ensuremath{\mathbf{X}}}
\newcommand{\bY}{\ensuremath{\mathbf{Y}}}
\newcommand{\bZ}{\ensuremath{\mathbf{Z}}}
\newcommand{\bG}{\ensuremath{\mathbf{G}}}
\newcommand{\bT}{\ensuremath{\mathbf{T}}}
\newcommand{\fF}{\ensuremath{\widetilde{F}}}
\newcommand{\FD}{\ensuremath{\mathsf{FD}}}
\newcommand{\Def}{\ensuremath{\mathsf{Fun}}}

\newcommand{\bpsi}{\ensuremath{\mathbf{\Psi}}}

\newcommand{\PH}{\ensuremath{\mathsf{PH}}}

\newcommand{\BFS}{\ensuremath{\mathsf{BFnS}}}

\newcommand{\DNNF}{\textsf{DNNF}}
\newcommand{\CNF}{\ensuremath{\mathsf{CNF}}}
\newcommand{\BDD}{\textsf{ROBDD/FBDD}}

\newcommand{\lits}[1]{\ensuremath{lits({#1})}}
\newcommand{\atoms}[1]{\ensuremath{atoms({#1})}}
\newcommand{\AIG}{\ensuremath{\mathsf{AIG}}}

\usetikzlibrary{decorations.pathreplacing} 
\usetikzlibrary{patterns}

\usepackage[ruled,vlined,linesnumbered,lined]{algorithm2e}

\tikzset{
block/.style={
  draw, 
  rectangle, 
  minimum height=0cm, 
  minimum width=2cm, align=center,
  }, 
line/.style={->,>=latex'}
}
\tikzstyle{galivenode}=[circle,fill=green!50!black,thick,inner sep=1pt,minimum size=4mm]
\tikzstyle{ralivenode}=[circle,fill=red!70!black,thick,inner sep=1pt,minimum size=4mm]

\tikzstyle{alivenode}=[circle,fill=black!80,thick,inner sep=1pt,minimum size=4mm]
\tikzstyle{deadnode}=[circle,fill=black!10,thick,inner sep=0pt,minimum size=4mm]
\tikzstyle{rdnode}=[circle,draw,fill=red!10,thick,inner sep=2pt,minimum size=4mm]
\tikzstyle{grnode}=[circle,draw,fill=green!10,thick,inner sep=2pt,minimum size=4mm]

\tikzstyle{ynode}=[rectangle,draw=black,fill=yellow!10,thick,inner sep=3pt,minimum size=4mm]
\tikzstyle{gnode}=[rectangle,fill=green!10,thick,inner sep=1pt,minimum size=4mm]
\tikzstyle{bnode}=[rectangle,fill=red!10,thick,inner sep=1pt,minimum size=4mm]

\tikzstyle{block} = [rectangle, draw, fill=blue!20, 
    text width=5.7em, text centered, rounded corners, minimum height=2cm]
\tikzstyle{line} = [draw, -latex']
\tikzstyle{cloud} = [draw, ellipse,fill=red!20, minimum size=4mm]

\newcommand{\bigO}{\mathcal{O}}
\newcommand{\NNF}{\ensuremath{\mathsf{NNF}}}
\newcommand{\wDNNF}{\textsf{wDNNF}}

\newcommand{\dDNNF}{\textsf{dDNNF}}
\newcommand{\FBDD}{\textsf{FBDD}}
\newcommand{\ROBDD}{\textsf{ROBDD}}

\newcommand{\syn}{\textsf{SynNNF}}

\newcommand{\SynNNF}{\syn}

\newcommand{\decomp}[2]{[\nnf{#2}]_{#1}}

\newcommand{\ol}[1]{\ensuremath{\overline{{#1}}}}
\newcommand{\nnf}[1]{\ensuremath{\widehat{{#1}}}}

\newcommand{\proj}[2]{\ensuremath{{#1}\!\!\downarrow\!\!{\small{#2}}}}
\newcommand{\suprt}[1]{\ensuremath{\mathsf{sup}({#1})}}

\newcommand{\xor}{\oplus}
\newcommand{\fbdd}{\ensuremath{\mathsf{FBDD}}}
\newcommand{\obdd}{\ensuremath{\mathsf{OBDD}}}
\newcommand{\robdd}{\ensuremath{\mathsf{ROBDD}}}

\newcommand{\rep}{\mathsf{rep}}
\newcommand{\pp}{\mathsf{P}}
\newcommand{\np}{\mathsf{NP}}

\newcommand{\vnp}{\mathsf{VNP}}
\newcommand{\perm}{\mathsf{perm}}
\newcommand{\pma}{\mathsf{pm}}
\newcommand{\uF}{\ensuremath{\mathsf{\widetilde{F}}}}
\newcommand{\uG}{\ensuremath{\mathsf{\widetilde{G}}}}

\newcommand{\CToSyn}{\ensuremath{\mathsf{C2Syn}}}
\newcommand{\form}[1]{\ensuremath{\mathsf{form}({#1})}}

\newcommand{\GASCK}{$\mathsf{GACKS}{}$}

\begin{document}

\title{Knowledge Compilation for Boolean Functional Synthesis}
\author{\IEEEauthorblockN{S. Akshay, Jatin Arora, Supratik Chakraborty, S. Krishna, Divya Raghunathan and Shetal Shah}
  \IEEEauthorblockA{Indian Institute of Technology Bombay, Mumbai, India}}

\maketitle

\begin{abstract}
Given a Boolean formula $F(\bX, \bY)$, where $\bX$ is a vector of
outputs and $\bY$ is a vector of inputs, the Boolean functional
synthesis problem requires us to compute a Skolem function vector
$\mathbf{\Psi}(\bY)$ for $\bX$ such that $F(\mathbf{\Psi}(\bY),\bY)$
holds whenever
$\exists \bX\,F(\bX,\bY)$ holds. 
In this paper, we investigate the relation between the representation
of the specification $F(\bX, \bY)$ and the complexity of synthesis. We
introduce a new normal form for Boolean formulas, called {\syn}, that
guarantees polynomial-time synthesis and also polynomial-time
existential quantification for some order of quantification of
variables.  We show that several normal forms studied in the knowledge
compilation literature are subsumed by {\syn}, although {\syn} can be
super-polynomially more succinct than them.  Motivated by these
results, we propose an algorithm to convert a specification in {\CNF}
to {\syn}, with the intent of solving the Boolean functional synthesis
problem. Experiments with a prototype implementation show that this
approach solves several benchmarks beyond the reach of
state-of-the-art tools.

\end{abstract}


\section{Introduction}
\label{sec:intro}
\emph{Boolean functional synthesis} is the problem
of synthesizing outputs as Boolean functions of inputs, while
satisfying a declarative relational specification between inputs and
outputs.  Also called \emph{Skolem function synthesis}, this problem
has numerous applications including certified QBF solving, reactive
control synthesis, circuit and program repair and the like.  While
variants of the problem have been studied since
long~\cite{lowenheim1910,boole1847}, there has been significant recent
interest in designing practically efficient algorithms for
Boolean functional synthesis. The resulting breed of
algorithms~\cite{fmcad2015:skolem,Rabe15,Rabe16,rsynth,rsynth:fmcad2017,bierre,jiang2,tacas2017,cav18,KMPS10,CFTV18,RTRS18}
have been empirically shown to work well on large collections of
benchmarks.  Nevertheless, there are not-so-large examples that are
currently not solvable within reasonable resources by any known
algorithm.  To make matters worse, it is not even fully understood
what properties of a Boolean relational specification or of its
representation make it amenable to efficient synthesis.  In this
paper, we take a step towards answering this question.  Specifically,
we propose a new sub-class of negation normal form called {\syn}, such
that every Boolean relational specification in {\syn} admits
polynomial-time synthesis.  Furthermore, a Boolean relational
specification admits polynomial-time synthesis (by any algorithm)
\emph{if and only if} there exists a polynomial-sized
\emph{refinement} of the specification in {\syn}.

To illustrate the hardness of Boolean functional synthesis, consider
the specification $F(\bX_1,\bX_2,\bY) \equiv (\bY =
(\bX_1 \times_{[n]} \bX_2)) \wedge (\bX_1 \neq 0\cdots01) \wedge
(\bX_2 \neq 0\cdots01)$, where $|\bY| = 2n$, $|\bX_1| = |\bX_2| = n$
and $\times_{[n]}$ denotes multiplication of $n$-bit unsigned
integers.  This specification asserts that $\bY$, viewed as a $2n$-bit
unsigned integer, is the product of $\bX_1$ and $\bX_2$, each viewed
as an $n$-bit unsigned integer different from $1$. The specification
$F(\bX_1, \bX_2, \bY)$ can be easily represented as a circuit of AND,
OR, NOT gates with $\bigO(n^2)$ gates. However, synthesizing $\bX_1$
and $\bX_2$ as functions of $\bY$ requires us to obtain a circuit that
factorizes a $2n$-bit unsigned integer into factors different from $1$,
whenever possible.  It is a long-standing open question whether such a
circuit of size polynomial in $n$ exists. Thus, although the
relational specification is succinctly representable, the outputs
expressed as functions of the inputs may not have any known succinct
representation.

It was recently shown~\cite{cav18} that unless some long-standing
complexity theoretic conjectures are falsified, Boolean functional
synthesis must necessarily require super-polynomial (or even
exponential) space and time. In the same work~\cite{cav18}, it was also shown that if a specification is represented in \emph{weak decomposable negation
normal form {\wDNNF}}, synthesis can be accomplished in time
polynomial in the size of the specification.  While this was a first
step towards identifying a normal form with the explicit objective of
polynomial-time synthesis, experimental results in~\cite{cav18}
indicate that {\wDNNF} doesn't really characterize specifications that
admit efficient synthesis. Specifically, experiments in~\cite{cav18} showed that a polynomial-time algorithm intended for synthesis from {\wDNNF} specifications ends up solving the synthesis problem for a large class of
specifications \emph{not in {\wDNNF}}.  This motivates us to ask if
there exists a weaker (than {\wDNNF}) sub-class of Boolean relational
specifications that admit polynomial-time synthesis.

\emph{We answer the above question affirmatively
in this paper, the polynomial dependence being quadratic in the number of
outputs and the size of the specification}. En route, we also show
that the weaker normal form, viz. {\syn}, admits polynomial-time
existential quantifier elimination of a set of variables for
\emph{some (not all)} order of quantification of variables. 
Applications of such quantifier elimination abound in
practice, viz. image computation in symbolic model checking, synthesis
of QBF certificates, computation of interpolants etc.  Note that
ensuring efficient quantifier elimination \emph{for some ordering} of
variables is simpler than ensuring efficient quantifier
elimination \emph{for all orderings} of variables -- the latter
having been addressed by normal forms like
{\DNNF}~\cite{darwiche-jacm}.

Our primary contributions can be summarized as follows:
\begin{itemize}
\item We present a new sub-class of negation normal form,
called {\syn}, that admits polynomial-time synthesis and 
quantifier elimination for a set of variables.
\item We show that {\SynNNF} is super-polynomially (in some cases,
exponentially) more succinct than several other sub-classes studied in
the literature (viz. {\wDNNF}, {\dDNNF}, {\DNNF}, {\FBDD}, {\ROBDD}),
unless some long-standing complexity theoretic conjectures are
falsified.
\item We show that by suitably weakening {\syn}, we can 
precisely characterize the class of Boolean specifications that admit
polynomial-time synthesis by a simple algorithm originally proposed
in~\cite{cav18}.
\item We define a natural notion of
 \emph{refinement of specifications w.r.t synthesis} and show that
every specification that admits polynomial-time synthesis necessarily
has a polynomial-sized refinement that is in {\syn}.
\item We present a novel
algorithm for compiling a Boolean relational specification in {\CNF}
to a \emph{refined} specification in {\syn}. We call
this \emph{knowledge compilation for synthesis and quantifier
elimination}.
\item Finally, we present experimental results that show that
synthesis by compiling to {\syn} solves a large set of benchmarks,
including several benchmarks beyond the reach of existing tools.
\end{itemize}

\noindent \emph{Related Work:}
The literature on knowledge compilation of Boolean functions is rich
and extensive~\cite{CD97,darwiche-jacm,dsharp,DM11}. While existential quantification or \emph{forgetting} of propositions
has been studied in~\cite{LLM11,DM11}, neither Boolean functional
synthesis nor existential quantification for \emph{some (not all)}
ordering of variables has received attention in earlier work on
knowledge compilation.  Sub-classes of negation normal forms like
{\DNNF} and other variants~\cite{DM11} admit efficient existential
quantification \emph{for all} orders in which variables are
quantified.  However, if we are interested in only the result of
existentially quantifying a given set of variables, these forms can be
unnecessarily restrictive and exponentially larger.  Recent work on
Boolean functional
synthesis~\cite{jiang2,fmcad2015:skolem,bierre,RTRS18,rsynth,tacas2017,cav18,bafsyn:fmcad2018}
has focused more on algorithms to directly synthesize outputs as
functions of inputs.  Some of these algorithms (viz.~\cite{rsynth,cav18,bafsyn:fmcad2018})
exploit properties of specific input representations for optimizing
the synthesis process. This has led to the articulation
of \emph{sufficient} conditions on representation of specifications
for efficient synthesis.  For example, \cite{KMPS10} suggested using
input-first {\ROBDD}s for efficient synthesis, and a quadratic-time
algorithm for synthesis from input-first {\ROBDD}s was presented
in~\cite{rsynth}.  This result was subsequently generalized
in~\cite{cav18}, where it was shown that specifications in {\wDNNF}
(which strictly subsumes {\ROBDD}s) suffice to give a quadratic-time
algorithm for synthesis.  As we show later, {\wDNNF} can
itself be generalized to {\syn}. In another line of investigation, it was shown~\cite{bafsyn:fmcad2018}
that if a {\CNF} specification is decomposed into an \emph{input-part}
and an \emph{output-part}, then synthesis can be achieved in time
linear in the size of the {\CNF} specification and $k$, where $k$ is
the smaller of the count of \emph{maximal falsifiable subsets (MFS)}
of the input-part and the count of \emph{maximal satisfiable subsets
(MSS)} of the output-part.  However, this does not yield an algorithm
whose running time is polynomial in the size of the representation
of $F(\bX,\bY)$.

The paper is organized as follows. After preliminaries, we present the new normal form {\syn} and its properties in Section~\ref{sec:compiled}. In Section~\ref{sec:refine}, we introduce the idea of refinement, which allows us to simplify the specification. In Section~\ref{sec:dsharp}, we describe an algorithm to compile any function into our normal form, followed in Section~\ref{sec:expt} by experimental results, before ending with a conclusion. Proofs of lemmas and theorems are mostly deferred to the appendix.

\section{Preliminaries and notations}
\label{sec:prelim}
A Boolean formula $F(z_1, \ldots z_p)$ on $p$ variables is a mapping
$F: \{0, 1\}^p \rightarrow \{0,1\}$.  The set of variables $\{z_1,
\ldots z_p\}$ is called the \emph{support} of the formula, and denoted
$\suprt{F}$.  We normally use $\bZ$ to denote the sequence $(z_1,
\ldots z_p)$.  For notational convenience, we will also use $\bZ$ 
to denote a set of variables, when there is no confusion.
A \emph{satisfying assignment} or \emph{model} of $F$ is a mapping of
variables in $\suprt{F}$ to $\{0,1\}$ such that $F$ evaluates to $1$
under this assignment.  If $\pi$ is a model of $F$, we write
$\pi \models F$ and use $\pi(z_i)$ to denote the value assigned to
$z_i \in \suprt{F}$ by $\pi$.  If $\bZ'$ is a subsequence of $\bZ$, we
use $\proj{\pi}{\bZ'}$ to denote the projection of $\pi$ on $\bZ'$,
i.e. $(\pi({z'}_1), \ldots \pi({z'}_k))$, where $k = |\bZ'|$.  We use
$\form{\proj{\pi}{\bZ'}}$ to denote the conjunction of \emph{literals}
(i.e. variables or their negation) corresponding to
$\proj{\pi}{\bZ'}$.  For example, if $\pi$ assigns $1$ to $z_1, z_3$
and $0$ to $z_{2}, z_{4}$ and $\bZ' = (z_1, z_4)$, then
$\form{\proj{\pi}{\bZ'}} = z_1 \wedge \neg z_4$.

\subsubsection{Negation normal form ($\NNF$)} This is the class of Boolean formulas in which (i) the only operators used are conjunction ($\wedge$), disjunction ($\vee$) and negation
($\neg$), and (ii) negation is applied only to variables. Every
Boolean formula can be converted to a semantically equivalent {\NNF}
formula.  Moreover, this conversion can be done in linear time for
representations like {\AIG}s, {\ROBDD}s, Boolean circuits etc.

\subsubsection{Unate formulas}
Let $F|_{z_i = 0}$ (resp. $F|_{z_i = 1}$) denote the positive
(resp. negative) \emph{cofactor} of $F$ with respect to $z_i$. Then,
$F$ is \emph{positive unate} in $z_i \in \suprt{F}$ iff $F|_{z_i =
0} \Rightarrow F|_{z_i = 1}$. Similarly, $F$ is
\emph{negative unate} in $z_i$ iff $F|_{z_i =1} \Rightarrow F|_{z_i
= 0}$.  A \emph{literal} $\ell$ is said to be \emph{pure} in an {\NNF}
formula $F$ iff $F$ has at least one instance of $\ell$ but no instance of
$\neg \ell$. If $z_i$ (resp. $\neg z_i$) is
pure in $F$, then $F$ is positive (resp. negative) unate in $z_i$.

\subsubsection{Independent support and functionally defined variables}
A subsequence $\bZ'$ of $\bZ$ is said to be an \emph{independent
support} of $F$ iff every pair of satisfying assignments $\pi,\pi'$ of
$F$ that agree on the assignment of variables in $\bZ'$ also agree on
the assignment of all variables in $\bZ$.  Variables not in $\bZ'$ are
said to be \emph{functionally defined} by the independent
support. Effectively, the assignment of variables in $\bZ'$ uniquely
determine that of functionally defined variables, when satisfying $F$.
{\CNF} encodings of Boolean functions originally specified as
circuits, {\ROBDD}s, {\AIG}s etc. often use Tseitin
encoding~\cite{tseitin68}, which introduces a large number of
functionally defined variables. 

\subsubsection{Boolean functional synthesis}
Unless mentioned otherwise, we use $\bX = (x_1, \ldots x_n)$ to denote
a sequence of Boolean outputs, and $\bY = (y_1, \ldots y_m)$ to denote
a sequence of Boolean inputs.  The \emph{Boolean functional synthesis}
problem, henceforth denoted {\BFS}, asks: given a Boolean formula
$F(\bX, \bY)$ specifying a relation between inputs $\bY$ and outputs $\bX$, determine functions
$\bpsi = (\psi_1(\bY), \ldots \psi_n(\bY))$ such that $F(\bpsi, \bY)$
holds whenever $\exists \bX F(\bX, \bY)$ holds. Thus, $\forall \bY
(\exists \bX\, F(\bX, \bY) \Leftrightarrow \left.F(\bpsi, \bY)\right)$
must be a tautology.  The function $\psi_i$ is called
a \emph{Skolem function} for $x_i$ in $F$, and $\bpsi$ is called a \emph{Skolem function vector} for
$\bX$ in $F$.

For $1 {\le} i {\le} j {\le} n$, we use $\bX_{i}^j$ to denote the
subsequence $(x_i, x_{i+1}, \ldots x_j)$.  If $i \le k < j$, we
sometimes use $(\bX_{i}^k,\bX_{k+1}^j)$ interchangeably with $\bX_i^j$
for notational convenience.  Let $F^{(i-1)}(\bX_i^n, \bY)$ denote
$\exists \bX_1^{i-1} F(\bX_1^{i-1}, \bX_i^n, \bY)$. It has been argued
in~\cite{fmcad2015:skolem,rsynth,tacas2017,Jian} that 
the {\BFS} problem for $F(\bX,\bY)$ can be solved by first ordering
the outputs, say as $x_1 \prec x_2 \cdots \prec x_n$, and then
synthesizing a function $\psi_i(\bX_{i+1}^n, \bY) \equiv
F^{(i-1)}(\bX_{i}^n, \bY)[x_i \mapsto 1]$ for each $x_i$.
This ensures that
$F^{(i-1)}(\psi_i, \bX_{i+1}^n, \bY) \Leftrightarrow \exists x_i F^{(i-1)}(x_i,
\bX_{i+1}^n, \bY)$.   Once all such $\psi_i$'s are obtained, one can
substitute $\psi_{i+1}$ through $\psi_{n}$ for $x_{i+1}$ through $x_n$
respectively, in $\psi_i$ to obtain a Skolem function for $x_i$ as a
function of $\bY$. The primary problem of using this approach as-is
is the exponential blow-up incurred in the size of the Skolem functions.

\subsubsection{DAG representations}
For an {\NNF} formula $F$, its DAG representation is naturally
induced by the structure of $F$.  Specifically, if $F$ is simply a
literal $\ell$, its DAG representation is a leaf labeled $\ell$.  If
$F$ is $F_1 ~\mathsf{op}~ F_2$ where
$\mathsf{op} \in \{\vee,\wedge\}$, its DAG representation is a node
labeled $\mathsf{op}$ with two children, viz. the DAG representations
of $F_1$ and $F_2$.  W.l.o.g. we assume that a DAG representation of
$F$ is always in a \emph{simplified} form, where $t \wedge 1 $, $t\vee
0$, $t \wedge t$ and $t \vee t$ are replaced by $t$, $t\wedge 0$ is
replaced by 0 and $t\vee 1$ is replaced by $1$ for every node $t$.  We
use $|F|$ for the node count in the DAG representation of $F$.

{\fbdd} and {\robdd} are well-known representations of Boolean
formulas and we skip their definitions. We briefly recall the
definitions of {\DNNF}, {\dDNNF} and {\wDNNF} below. Let $\alpha$ be
the subformula represented by an internal node $N$ (labeled by
$\wedge$ or $\vee$) in a DAG representation of an NNF formula $F$. We
use $\lits{\alpha}$ to denote the set of literals labeling leaves that
have a path to the node $N$ representing $\alpha$ in the DAG
representation of $F$. We also use $\atoms{\alpha}$ to denote the
underlying set of variables in $\suprt{F}$ that appear in
$\lits{\alpha}$.  For each $\wedge$-labeled internal node $N$ in the
DAG of $\Ff$ with $\alpha = \alpha_1 \wedge \ldots \wedge \alpha_k$
being the subformula represented by $N$, if for all distinct indices
$r, s \in \{1, \ldots k\}$,
$\atoms{\alpha_r}\cap\atoms{\alpha_s}=\emptyset$, then $F$ is said to
be in \DNNF{}~\cite{darwiche-jacm}. If, instead, for all distinct
indices $r, s \in \{1, \ldots k\}$,
$\lits{\alpha_r} \cap \{\neg\ell \mid \ell \in \lits{\alpha_s}\}
= \emptyset$, then $F$ is said to be in \wDNNF{}~\cite{cav18}. Finally
$\Ff(\bX, \bY)$ is said to be
in \textsf{deterministic} \DNNF{}(or \dDNNF)~\cite{DM11} if $\Ff$ is
in $\DNNF$ and for each $\vee$-labeled internal node $D$ in the DAG of
$\Ff$ with $\beta = \beta_1 \vee \ldots \vee \beta_k$ being the
subformula represented by $D$, $\beta_r \wedge \beta_s$ is a
contradiction for all distinct indices $r, s$.

\subsubsection{Positive form of input specification}
Given a specification $\Ff(\bX, \bY)$ in {\NNF}, we denote by
$\nnf{\Ff}(\bX, \ol{\bX}, \bY)$ the formula obtained by replacing
every occurrence of $\neg x_i ~(x_i \in \bX)$ in $\Ff$ with a fresh
variable $\ol{x_i}$. This is also called the \emph{positive form} of
the specification and has been used earlier in~\cite{tacas2017}. 
Observe that for any $\Ff$ in {\NNF}, $\nnf{\Ff}$ is positive
unate (or \emph{monotone}) in all variables in $\bX$ and $\ol{\bX}$.
For $i \in \{1, \ldots n\}$, we sometimes split $\bX$ into two parts,
$\bX_{1}^i$ and $\bX_{i+1}^n$, and represent
$\nnf{F}(\bX, \ol{\bX}, \bY)$ as
$\nnf{F}(\bX_{1}^i,\bX_{i+1}^n, \ol{\bX}_{1}^i, \ol{\bX}_{i+1}^n, \bY)$. 
For $b, c \in \{0,1\}$,  let $\mathbf{b}^{i}$
 (resp. $\mathbf{c}^{i}$) denote a vector of $i$ $b$'s
 (resp. $c$'s). For notational convenience,
we use
$\nnf{F}(\mathbf{b}^i, \bX_{i+1}^n, \mathbf{c}^i, \ol{\bX}_{i+1}^n, \bY)$
to denote
$\nnf{F}(\bX_{1}^i, \bX_{i+1}^n, \ol{\bX}_{1}^i, \ol{\bX}_{i+1}^n, \bY)|_{\bX_{1}^i
= \mathbf{b}^i, \ol{\bX}_{1}^i = \mathbf{c}^i}$.

\section{A New Normal Form for Efficient Synthesis}
\label{sec:compiled}
In~\cite{cav18}, it was shown that if $F(\bX,\bY)$ is represented as a {\BDD} or in {\DNNF} or in {\wDNNF} form, Skolem functions can be synthesized in time polynomial in $|F|$.  In this section, we define a new normal form called {\syn}
that subsumes and is more succinct than these other normal forms, and yet
guarantees efficient synthesis of Skolem functions. 

\begin{definition}
Given a specification $F(\bX,\bY)$, for every $i \in \{1, \ldots n\}$ we define the \emph{$i^{th}$-reduct of $\nnf{\Ff}$}, denoted
$\decomp{i}{\Ff}$, to be 
$\nnf{\Ff}(1^{i-1},{\bX}_{i}^n,1^{i-1},\ol{\bX}_{i}^n,\bY)$. We also
define $\decomp{n+1}{\Ff}$ to be
$\nnf{\Ff}(1^n,1^n,\bY)$.
\end{definition} 
Note that $\decomp{1}{\Ff}$ is the same as $\nnf{F}$, and $\suprt{\decomp{i}{\Ff}} = \bX_i^n \cup \ol{\bX}_i^n \cup \bY$ for $i \in \{1, \ldots n\}$. 
\begin{example}
\label{eg:1} Consider the {\NNF} formula $K(x_1,x_2,y_1,y_2)=(x_1 \vee x_2) \wedge (\neg x_2 \vee y_1)\wedge (\neg y_1 \vee y_2)$. Then $\nnf{K}=((x_1\vee x_2)\wedge (\ol{x_2}\vee y_1)\wedge (\neg y_1 \vee y_2))$.
Thus, we  have $\decomp{1}{K}=\nnf{K}$ 
and $\decomp{2}{K}= \nnf{K}[x_1 \mapsto 1,\ol{x_1} \mapsto 1] = (\ol{x_2}\vee y_1) \wedge (\neg y_1 \vee y_2)$.
\end{example}

Next, we define a useful property for the $i^{th}$-reduct, which will be crucial for efficient synthesis of Skolem functions.
\begin{definition}
\label{def:alphadef}
Given $F(\bX,\bY)$, let  $\alpha_i^{jk}$ denote $\decomp{i}{F}[x_i \mapsto j, \ol{x}_i \mapsto k, \ol{\bX}_{i+1}^n \mapsto \neg \bX_{i+1}^n]$, where  $j, k \in \{0,1\}$.  We say that $\decomp{i}{F}$ is $\wedge_i$-unrealizable if $\zeta= \alpha_i^{11} \wedge \neg\alpha_i^{10} \wedge \neg\alpha_i^{01}$ is unsatisfiable.
\end{definition}
Intuitively, we wish to say that there is no assignment to $\bX_{i+1}^n$ and $\bY$ such that $\decomp{i}{\Ff}$ is equivalent to $x_i \wedge \overline{x}_i$.
The formula $\zeta$ captures this semantic condition. Indeed, if an assignment makes $\zeta$  true,  then it also makes $\decomp{i}{F}$ equivalent to $x_i \wedge  \ol{x}_i$ (i.e., $\decomp{i}{F}=1$ for $x_i,\ol{x_i}$ having values $(1,1)$, but not for $(0,1)$, $(1,0)$, $(0,0)$).  Note
that since $\decomp{i}{\Ff}$ is positive unate in $x_i$ and $\ol{x_i}$, $\zeta$ is satisfiable iff $\zeta \wedge  \neg\alpha_i^{00}$ is satisfiable; we need not conjoin $\neg\alpha_i^{00}$ in the definition of $\zeta$.

A sufficient condition for $\decomp{i}{\Ff}$ to be $\wedge_i$-unrealizable is that in the DAG representation of $\decomp{i}{\Ff}$, there is no pair of paths -- one from  $x_i$ and the other from $\ol{x_i}$ -- which meet for the first time at an $\wedge$-labeled node. In Example~\ref{eg:1}, $\decomp{1}{K}$ is $\wedge_1$-unrealizable  since there is no leaf labeled $\ol{x_1}$ in its DAG representation. Similarly, $\decomp{2}{K}=(\ol{x_2}\vee y_1)\wedge(\neg y_1 \vee y_2)$ is $\wedge_2$-unrealizable as there is no leaf labeled $x_2$ in the DAG representation of $\decomp{2}{K}$ (although such a leaf exists in the DAG representation of $\decomp{1}{K}$).

\begin{example}
\label{eg:2}
Let $H(x_1, x_2, y_1,y_2) =(x_1 \vee x_2 \vee y_1) \wedge (\neg x_1 \vee (\neg x_2 \wedge y_2))$. 
Then $\nnf{H}(\bX,\overline{\bX},\bY) = (x_1 \vee x_2 \vee y_1) \wedge (\ol{x_1} \vee (\ol{x_2} \wedge y_2))$.
Using the notation in Definition~\ref{def:alphadef}, $\alpha_1^{11}= 1$, 
$\alpha_1^{10}= \neg{x}_2 \wedge y_2$ 
and $\alpha_1^{01}=(x_2 \vee y_1)$. 
There is an assignment ($x_2=0,y_2=0,y_1=0)$ such that  $(\alpha_1^{11} \wedge \neg\alpha_1^{10} \wedge \neg\alpha_1^{01})$ is satisfiable. Hence $\decomp{1}{H}$ is not $\wedge_1$-unrealizable (equivalently, it is $\wedge_1$-realizable). However, $\decomp{2}{H}= \nnf{H}[x_1\mapsto 1,\ol{x_1} \mapsto 1] = 1$;
hence it is vacuously $\wedge_2$-unrealizable.
\end{example}
 \begin{definition}
A formula $\Ff(\bX, \bY)$ is said to be in \textsf{synthesizable} $\NNF$ (or \syn~) wrt the sequence $\bX$ if $\Ff$ is in $\NNF$, and for all $1\leq i\leq n$, $\decomp{i}{\Ff}$ is $\wedge_i$-unrealizable.
\end{definition}

\noindent In Examples~\ref{eg:1}, \ref{eg:2}, $K$ is in $\syn$, while $H$ is not. Also neither of them are in $\DNNF$ or $\wDNNF$. Additionally, the functions as presented do not correspond to $\BDD$ representations either. We now show \emph{three} important properties of {\syn} which motivate our proposal of {\syn} as a normal form for synthesis and existential quantification.
\subsubsection{\syn{} leads to efficient quantification and synthesis}
Our first result is that existentially quantifying $\bX$ and synthesizing $\bX$ are easy for \syn.
\begin{theorem}
\label{thm:synnnf-exists-synth}
Suppose $\Ff(\bX,\bY)$ is in {\syn}. Then,
\begin{enumerate}[(i)]
\item \label{thm:synnnf-exists}
$\exists \bX_1^i F(\bX,\bY) \Leftrightarrow \decomp{i+1}{F}[\ol{\bX}_{i+1}^n \mapsto \neg\bX_{i+1}^n]$ for $i\in\{1, \ldots, n\}$,
\item \label{thm:synnnf-synth}
Skolem function vector $\Psi_1^n$ for $\bX_1^n$ can be computed in $\bigO(n^2\cdot |F|)$ time and $\bigO(n\cdot |F|)$ space, where $|\bX| = n$.
\end{enumerate}
\end{theorem}
\begin{proof} 
The proof of Part (i) is similar to that of Theorem~2(a) in~\cite{cav18}, and follows by induction on $i$. For $i=1$, $\exists \bX_1^1 F(\bX,\bY)\Leftrightarrow \nnf{F}(1,\bX_2^n,0,\neg \bX_2^n,\bY)\vee \nnf{F}(0,\bX_2^n,1,\neg\bX_2^n,\bY) \implies \nnf{F}(1,\bX_2^n,1,\neg\bX_2^n,\bY)=\decomp{2}{F}[\ol{\bX}_2^n\mapsto \neg \bX_2^n]$ (by positive unateness of $\nnf{F}$ in $x_1,\ol{x_1}$).
Conversely, as $F$ is in \syn, $\decomp{2}{F}$ is $\wedge_2$-unrealizable, which implies that with notation as in Definition~\ref{def:alphadef}, $\alpha_1^{11}\implies \alpha_1^{10}\vee \alpha_1^{01}$, i.e., $\nnf{F}(1,\bX_2^n,1,\neg\bX_2^n,\bY)\implies \nnf{F}(1,\bX_2^n,0,\neg \bX_2^n,\bY)\vee \nnf{F}(0,\bX_2^n,1,\neg\bX_2^n,\bY)$.  This give us the proof in the reverse direction, i.e., $\decomp{2}{F}[\ol{\bX}_2^n\mapsto \neg \bX_2^n]\implies \exists \bX_1^1 F(\bX,\bY)$.

Suppose the statement holds for $1 \le i <n$. We will show that it holds for $i+1$ as well. By inductive hypothesis and definition of existential quantification, $\exists \bX_1^{i+1} F(\bX,\bY)\Leftrightarrow \exists x_{i+1} \decomp{i+1}{F}[\ol{\bX}_{i+1}^n \mapsto \neg\bX_{i+1}^n]\Leftrightarrow \decomp{i+1}{F}[x_i\mapsto 1,\ol{\bX}_{i+1}^n \mapsto \neg\bX_{i+1}^n]\vee \decomp{i+1}{F}[x_i\mapsto 0,\ol{\bX}_{i+1}^n \mapsto \neg\bX_{i+1}^n]$. Again, using unateness of $\decomp{i+1}{F}$
in $x_{i+1}$ and $\ol{x_{i+1}}$ in one direction, and using the defining property of {\syn} ($\alpha_{i+1}^{11}\implies \alpha_{i+1}^{10}\vee \alpha_{i+1}^{01}$) in the other direction, we obtain $\exists \bX_1^{i+1} F(\bX,\bY) \Leftrightarrow
\decomp{i+2}{F}[\ol{\bX}_{i+2}^n\mapsto \neg \bX_{i+2}^n]$.

Part(ii): For $i \in \{1, \ldots n\}$, let
$\psi'_i(\bX_{i+1}^n,\bY)$ denote $\decomp{i}{F}[x_i\mapsto 1,\ol{x}_i\mapsto
0,\ol{\bX}_{i+1}^n \mapsto \neg \bX_{i+1}^n]= \alpha_i^{10}$. Further,
from $n$ to $1$, we recursively define $\psi_n(\bY)=\psi'_n(\bY)$ and
$\psi_i(\bY)=\psi'_i({\Psi}_{i+1}^n(\bY),\bY)$.
We can now show that $\psi_i(\bY)$ is indeed a correct Skolem function
for $x_i$ in $F$. Starting from $n$ to $1$, we know from the
preliminaries that $F^{(n-1)}[x_n\mapsto 1]$ gives a correct Skolem
function for $x_n$ in $F$.  From part (i) above,
$F^{(n-1)} \Leftrightarrow \decomp{n}{F}[\ol{\bX_n^n}\mapsto \neg
\bX_n^n]$. Hence $\alpha_n^{10} = \psi_n = \psi'_n$ gives a
correct Skolem function for $x_n$ in $F$.  
For any
$i \in \{1,\ldots n-1\}$, assuming that $\Psi_{i+1}^n$ gives a correct Skolem
function vector for $\bX_{i+1}^n$ in $F$, the same argument shows that
$\psi'_i({\psi}_{i+1}^n(\bY),\bY)$ is a correct Skolem function
for $x_i$ in $F$.

Finally, note that $|\psi_n|$ is at most $|\nnf{F}|$, which is in
$\bigO(|F|)$.  A DAG representation of $\psi_{n-k}$ requires a fresh
copy of $\decomp{n-k}{F}$, but can re-use the DAG representations of
$\psi_{j}$ for $j \in \{n-k+1, \ldots n\}$ as sub-DAGs.  Thus,
$|\psi_{n-k}|$ is in $\bigO(k\cdot |F|)$. Hence, if we use a
multi-rooted DAG to represent all Skolem functions together, we need
only $\bigO(n\cdot|F|)$ nodes. The time required is in
$\bigO(n^2\cdot|F|)$ since the resulting DAG has $\sum_{k=1}^n k$
edges (root of $\psi_j$ connects to a leaf of every $\psi_i$ for $i <
j$).
\end{proof}
The above polynomial-time strategy based on $\decomp{i}{F}$ was used in~\cite{cav18} for computing over-approximations of Skolem functions $\psi_i(\bX_{i+1}, \bY)$ for each $x_i \in \bX$.  Specifically, it was shown that $\decomp{i}{F}[x_i \mapsto 1, \ol{x_i} \mapsto 1]$ over-approximates $\exists \bX_1^i F(\bX,\bY)$ and $\decomp{i}{F}[x_i \mapsto 1, \ol{x_i}\mapsto 0]$ over-approximates a Skolem function for $x_i$ in $F$.  In the remainder of this paper, we refer to the functions $\psi_i$ used in the proof of Part (ii) above as {\GASCK} functions (after the author names of~\cite{cav18}).  We use $\Psi_1^n$ to denote the \GASCK\ (Skolem) function vector  $(\psi_1,\ldots,\psi_n)$.

\subsubsection{Succinctness of \syn} {\syn} strictly subsumes many known representations used for efficient analysis of Boolean functions. In the following theorem, sizes and times are in terms of the number of input and output variables. 
\begin{theorem}
\label{thm:succinctness}
\begin{enumerate}[(i)]
\item \label{thm:succinct-other-dds} Every specification in \BDD, \dDNNF{}, \DNNF{} or \wDNNF{} form is either already in \syn{} or can be compiled in linear time to \syn.
\item There exist poly-sized \syn{} specifications that only admit
\begin{enumerate}
\item exponential sized \FBDD{} representations.
\item super-polynomial sized \dDNNF{} representations, unless $\pp=\vnp$
\item super-polynomial sized \wDNNF{} and \DNNF{} representations, unless $\pp=\np$.
\end{enumerate}
\item There exist poly-sized NNF-representations that only admit super-polynomial sized \syn{} representations, unless the polynomial hierarchy collapses.
\end{enumerate}
\end{theorem}
In the above, $\vnp$ is the algebraic analogue of $\np$~\cite{Valiant79}. Also, (iii) shows that we cannot always hope to obtain a
succinct {\syn} representation. 

\subsubsection{\syn{} ``almost'' characterizes efficient synthesis using \GASCK~ functions}
\label{sec:almost-char}
We now show that {\syn} precisely characterizes specifications that admit linear-time existential quantification of output variables strengthening Theorem~\ref{thm:synnnf-exists-synth}(\ref{thm:synnnf-exists}). Further, a slight weakening of $\SynNNF$ condition by restricting assignments on $\bX_{i+1}^n$ gives us a necessary and sufficient condition for poly-time synthesis using {\GASCK} functions.

\begin{restatable}{theorem}{characmain}
\label{thm:characmain}
Given a relational specification $F(\bX,\bY)$,
\begin{enumerate}[(i)]
\item \label{thm:existscharac} $F$ is in \syn\ iff 
$\exists \bX_1^i F(\bX,\bY) \Leftrightarrow \decomp{i+1}{F}[\ol{\bX}_{i+1}^n \mapsto \neg\bX_{i+1}^n]$
\item  \label{thm:skfunchar} The \GASCK-function vector $\Psi_1^n$ is a Skolem function vector for $\bX_1^n$ in $F(\bX, \bY)$
iff $\decomp{i}{\Ff}[\bX_{i+1}^n \mapsto \Psi_{i+1}^n, \ol{\bX_{i+1}^n}\mapsto \neg \Psi_{i+1}^n]$ is $\wedge_i$-unrealizable  for all $i \in \{1 \ldots n\}$.
\end{enumerate}
\end{restatable}
In~\cite{fmcad2015:skolem}, it was shown that an \emph{error formula} $\varepsilon$ for $\Psi_1^n$, defined as $F(\bX,\bY) \wedge \neg F(\bX',\bY) \wedge \bigwedge_{i=1}^n(x_i' \leftrightarrow \Psi_i)$ is unsatisfiable iff $\Psi_1^n$ is a Skolem function vector for $F$.
Therefore, an (un)satisfiability check for $\varepsilon$ serves to
check if $\decomp{i}{\Ff}[\bX_{i+1}^n \mapsto \Psi_{i+1}^n]$ is
$\wedge_i$-unrealizable for all $i \in \{1 \ldots n\}$.  Further, in~\cite{cav18}, it was observed experimentally, that \GASCK~functions give correct Skolem functions, even when the specifications are not in {\wDNNF}. This surprising behavior, which was left unexplained in~\cite{cav18}, can now be explained using {\syn}, thanks to Theorem~\ref{thm:characmain}(\ref{thm:skfunchar}).

Note that Theorem~\ref{thm:characmain}(\ref{thm:skfunchar}) weakens
the requirement of {\syn} since $\bX_{i+1}^n$ are constrained to take
only the values defined by $\Psi_{i+1}^n$.  For an example of a
specification not in \syn~ for which \GASCK~ functions are correct Skolem
functions, consider again $H$ from Example~\ref{eg:2}, which we saw
was not in \syn. 
In this case,
$\psi'_1(x_2,\bY)=\decomp{1}{H}[x_1 \mapsto 1, \overline{x}_1 \mapsto
0, \ol{x}_2 \mapsto \neg x_2]=\neg x_2 \wedge y_2$ and $\psi_2(\bY)
= \psi'_2(\bY)=\decomp{2}{H}[x_2 \mapsto 1, \overline{x}_2 \mapsto
0]=1$. Therefore, $\psi_1(\bY)=\psi'_1[x_2 \mapsto \psi_2(\bY)]=0$. It can be verified that
$x_1=\psi_1(\bY) = 0, x_2=\psi_2(\bY) = 1$ is indeed a correct Skolem
function vector for $\bX$ in $H$.  Also, $H$ satisfies the condition of
Theorem~\ref{thm:characmain}(\ref{thm:skfunchar}) since
$\decomp{1}{H}[x_2 \mapsto \psi_2, \overline{x}_2 \mapsto \neg \psi_2]=\ol{x}_1 \notiff (x_1 \wedge \ol{x}_1)$, and $\decomp{2}{H}=1$.

\section{Refinement for Synthesis}
\label{sec:refine}
Given a specification $F(\bX,\bY)$, sometimes it is easier to solve
the {\BFS} problem for a ``simpler'' specification $\uF(\bX,\bY)$ 
such that a solution for $\uF$ also serves as a solution for $F$.
While ``simplifications'' of this nature have been used in earlier
work~\cite{fmcad2015:skolem,cav18,Rabe16,CFTV18}, we formalize this notion
below as one of refinement.
\begin{definition}
  \label{def:refine}
  Let $F(\bX,\bY)$ and $\uF(\bX,\bY)$ be Boolean relational
  specifications on the same input and output vectors. We say that
  \emph{$\uF$ refines $F$ w.r.t. synthesis}, denoted $\uF \preceq_{syn}
  F$, iff the following conditions hold:
  (a) $\forall \bY \left(\exists \bX F(\bX, \bY) \Rightarrow \exists \bX' {\uF}(\bX', \bY))\right)$, and
  (b) $\forall \bY \forall \bX' \left(\left(\exists \bX F(\bX, \bY) \wedge {\uF}(\bX',\bY)\right)\Rightarrow F(\bX', \bY)\right)$.
\end{definition}
Informally, condition (a) specifies that $\uF$ doesn't restrict the
set of input valuations (i.e. $\bY$) over which the specification $F$
can be satisfied, and condition (b) specifies that for all such input
valuations $\bY$, any $\bX'$ that satisfies $\uF$ also satisfies
$F$.  
\begin{restatable}{lemma}{skrefines}
If $\uF \preceq_{syn} F$, every Skolem function vector for
$\bX$ in $\uF$ is also a Skolem function vector for $\bX$ in $F$.
\end{restatable}
We say $\uF$ \emph{refines} $F$ w.r.t. synthesis because the set of
all Skolem function vectors for $\bX$ in $\uF$ is a subset of that for
$\bX$ in $F$.  Note that Definition~\ref{def:refine} provides a direct
2QBF-SAT based check of whether $\uF$ refines $F$ without referring to
the details of how $\uF$ is obtained from $F$.




\begin{example}
  \label{ex:ref1}
Let $G(x_1, x_2, y_1, y_2) \equiv (\neg x_1 \vee x_2 \vee y_1) \wedge
(x_1 \vee \neg x_2)\wedge (x_1 \vee \neg y_1)\wedge (x_2 \vee y_2)$
and $\uG(x_1, x_2, y_1, y_2) \equiv x_2 \wedge x_1$.  Although $G
\not\Leftrightarrow \uG$, both conditions (a) and (b) of
Definition~\ref{def:refine} are satisfied; hence $\uG \preceq_{syn} G$.
\end{example}
\noindent The following are easy consequences of Definition~\ref{def:refine}.
\begin{restatable}{proposition}{refineprops}
  \label{prop:refine-props}
  \begin{enumerate}
  \item \label{prop:a} $\preceq_{syn}$ is a reflexive and transitive
    relation on all Boolean relational specifications on $\bX \cup \bY$.
  \item \label{prop:aa} If $\bigwedge_{y_j \in \bY}\left(F|_{y_j=0}
    \Leftrightarrow F|_{y_j=1}\right)$ and $\pi \models F(\bX,\bY)$,
    then $\form{\proj{\pi}{\bX}} \preceq_{syn} F$.
  \item \label{prop:aaa} If $\bigwedge_{x_i \in \bX}\left(F|_{x_i=0}
    \Leftrightarrow F|_{x_i=1}\right)$,
    then $1 \preceq_{syn} F$.
  \item \label{prop:b} If $F$ is positive (resp. negative) unate in
    $x_i \in \bX$, then $x_i \wedge F|_{x_i = 1}$ (resp. $\neg x_i
    \wedge F|_{x_i = 0}$) $\preceq_{syn} F$.
  \item \label{prop:c}
    If $\uF_1 \preceq_{syn} F_1$ and $\uF_2 \preceq_{syn} F_2$, then
    \begin{enumerate}
    \item $(\uF_1 \vee \uF_2) \preceq_{syn} (F_1 \vee F_2)$.
    \item $(\uF_1 \wedge \uF_2) \preceq_{syn} (F_1 \wedge F_2)$ if
      the output supports of $F_1$ and $F_2$, and similarly of $\uF_1$ and $\uF_2$, are disjoint.
    \end{enumerate}
  \end{enumerate}       
\end{restatable}                      
Propositions~\ref{prop:refine-props}(\ref{prop:aa}) and
\ref{prop:refine-props}(\ref{prop:aaa}) effectively require
$F(\bX,\bY)$ to be semantically (but not necessarily syntactically)
independent of $\bY$ and $\bX$ respectively.  While these may appear
to be degenerate cases, we will soon see that both these propositions
turn out to be useful when recursively compiling a {\CNF}
specification into refined {\syn} specification. Interestingly, a
version of Proposition~\ref{prop:refine-props}(\ref{prop:b}) was used
in a pre-processing step of BFSS~\cite{cav18}, although the precise
notion of refinement w.r.t.  synthesis was not defined there. Thanks
to Definition~\ref{def:refine}, we can now generalize
Proposition~\ref{prop:refine-props}(\ref{prop:b}) to refine a
specification even when $F$ is not unate in any output variable.
We discuss below how this can be done.

Suppose the specification $F(\bX,\bY)$ uniquely defines an output
variable as a function of other input and output variables.  For
example, if $F(\bX,\bY) \equiv (\neg x_i \vee x_j) \wedge (\neg
x_i \vee y_k) \wedge (x_i \vee \neg x_j \vee \neg y_k) \wedge \cdots$,
then $F(\bX,\bY) \Rightarrow \left(x_i \Leftrightarrow (x_j \wedge
y_k)\right)$.  Such specifications arise naturally when a non-{\CNF}
Boolean formula is converted to {\CNF} via Tseitin
encoding~\cite{tseitin68}.  Variables like $x_i$ above are said to be
\emph{functionally determined} (henceforth called {\FD}) in $F$,
and implied functional
dependencies like $\left(x_i \leftrightarrow (x_j \wedge y_k)\right)$
are called \emph{functional definitions} (henceforth called
\emph{\fdefs}) of {\FD} variables in $F$.

Let $\bT \subseteq \bX$ be a set of {\FD} output variables in $F$, and
let ${\Def}_{\bT}$ be the conjunction of {\fdefs} of all
variables in $\bT$.  We say that $(\bT, {\Def}_{\bT})$ is
an \emph{acyclic system of {\fdefs}} if no variable in $\bT$
transitively depends on itself via the functional definitions in
${\Def}_{\bT}$.  In other words, ${\Def}_{\bT}$ induces an acyclic
system of functional dependencies between variables in $\bT$.  For
$x_i \in \bX \setminus {\bT}$, define $\theta_{F,\bT,x_i,a}$ to be
the formula \scriptsize $\left(F(\bX, \bY)|_{x_i=a} \wedge
\bigwedge_{x_j \in \bX \setminus (\bT \cup \{x_i\})}(x_j
\Leftrightarrow x_j')\right.$ $\wedge$
$\left. {\Def}_{\bT}(\bX',\bY)|_{x_i' = 1-a}\right)$ $\Rightarrow
F(\bX', \bY)|_{x_i' = 1-a}$ \normalsize , where $a \in \{0,1\}$ and
$\bX'$ is a sequence of fresh variables $(x_1', \ldots x_n')$.
Informally, $\theta_{F,\bT,x_i,a}$ asserts that if the specification
$F$ can be satisfied by setting a non-{\FD} output $x_i$ to $a$, then
it can also be satisfied by setting $x_i$ to the complement value
($1-a$), while preserving the values of all other non-{\FD} outputs.
The {\FD} outputs in ${\bT}$ must of course be set as per the
functional definitions in ${\Def}_{\bT}$.
\begin{restatable}{lemma}{refine}
\label{lem:refine-further} Let $(\bT, {\Def}_{\bT})$ be an
  acyclic system of {\fdefs} in $F$.
  \begin{enumerate}
  \item \label{lem:ref-further-1} If $\bX = \bT$, then ${\Def}_{\bT}
    \preceq_{syn} F$.
  \item \label{lem:ref-further-2} If $\bX \setminus \bT \neq \emptyset$,
    then for every $x_i \in \bX\setminus\bT$, we have:\\
    If $\theta_{F,\bT,x_i,0}$ is a tautology, then $(x_i
        \wedge F|_{x_i=1}) \preceq_{syn} F$.  Similarly, if
        $\theta_{F,\bT,x_i,1}$ is a tautology, then $(\neg x_i \wedge
        F|_{x_i = 0}) \preceq_{syn} F$.
  \end{enumerate}
  \end{restatable}
If $\bT = \emptyset$,
Lemma~\ref{lem:refine-further}(\ref{lem:ref-further-2}) simply reduces
to Proposition~\ref{prop:refine-props}(\ref{prop:b}).  However, if
$\bT \neq \emptyset$ (as is often the case),
Lemma~\ref{lem:refine-further}(\ref{lem:ref-further-2}) shows  that
$x_i\wedge F|_{x_i=1}$ (resp. $\neg x_i \wedge F|_{x_i=0}$) can refine $F$ even if $F$ is not
positive (resp. negative) unate in $x_i$.  As an illustration, the
specification $G(x_1, x_2, y_1, y_2)$ in Example~\ref{ex:ref1} is not
unate in either $x_1$ or $x_2$.
However, with $\bT = \{x_1\}$ and ${\Def}_{\bT} \equiv (x_1
\Leftrightarrow (x_2 \vee y_1))$, we have $\theta_{F,\bT,x_2,0} \equiv
1$. Hence, $x_2 \wedge G|_{x_2=1} \equiv (x_1 \wedge x_2)
\preceq_{syn} G$.  When $F$ is refined by an application of
Lemma~\ref{lem:refine-further}(\ref{lem:ref-further-2}), we say that
$F$ is refined by \emph{pivoting on} $x_i$.
  
\begin{restatable}{lemma}{refinesubsume}
  \label{lem:refine-subsume}
Let $(\bT, {\Def}_{\bT})$ and $(\bT', {\Def}_{\bT'})$ be acyclic
systems of {\fdefs} in $F$, where $\bT' \subseteq \bT \subseteq \bX$
and ${\Def}_{\bT} \equiv {\Def}_{\bT'} \wedge
{\Def}_{\bT\setminus\bT'}$.  For $a \in \{0,1\}$, if
$\theta_{F,\bT',x_i,a}$ is a tautology, then so is
$\theta_{F,\bT,x_i,a}$.
\end{restatable}
Lemma~\ref{lem:refine-subsume}, along with
Lemma~\ref{lem:refine-further}(\ref{lem:ref-further-2}), shows that
if $\bT' \subsetneq \bT \subseteq \bX$, the system of acyclic {\fdefs}
$(\bT, {\Def}_{\bT})$ potentially provides more opportunities for
refinement compared to $(\bT', {\Def}_{\bT'})$.  Hence, it is
advantageous to augment the set $\bT$ of {\FD} outputs (and
correspondingly ${\Def}_{\bT}$) whenever possible.

The following theorem suggests that compiling a given specification to
a refined {\syn} specification (as opposed to an equivalent {\syn}
specification) holds promise for Boolean functional synthesis. 
\begin{restatable}{theorem}{complete}
\label{thm:synnnf-complete}
For every relational specification $F(\bX,\bY)$, there exists a
polynomial-sized Skolem function vector for $\bX$ in $F$ iff there
exists a {\syn} specification $\uF(\bX,\bY)$ such that
$\uF\preceq_{syn} F$ and $\uF$ is polynomial-sized in $F$.
  \end{restatable}
Theorem~\ref{thm:synnnf-complete} guarantees that whenever a
polynomial-sized Skolem function vector exists for a specification
$F(\bX,\bY)$, there is also a polynomial-sized refined specification
in {\syn}.  It is therefore interesting to ask if we can compile
$F(\bX,\bY)$ to a ``small enough'' {\syn} specification $\uF(\bX,\bY)$
that refines $F$.  In the next two sections, we present such a
compilation algorithm and results of our preliminary experiments using
this algorithm.  Note that as shown in~\cite{cav18}, there exist
problem instances for which there are no polynomial-sized Skolem
function vectors, unless the Polynomial Hierarchy ({\PH}) collapses.
Thus, any algorithm for compilation to {\syn} must incur
super-polynomial blow-up (unless {\PH} collapses).  Nevertheless, as
our experiments show, the compilation-based approach works reasonably
well in practice, even solving benchmarks beyond the reach of existing
state-of-the-art {\BFS} tools.

\section{A Refining {\CNF} to {\SynNNF} Compiler}
\label{sec:dsharp}
We now describe {\CToSyn} -- an algorithm that takes as input a {\CNF}
specification $F(\bX,\bY)$ given as a set of clauses, and outputs a
DAG representation of a {\syn} specification $\fF(\bX,
\bY)$ that refines $F(\bX, \bY)$ w.r.t. synthesis.  Given a set
$\mathcal{S}$ of clauses, we use $\cnfform{\mathcal{S}}$ to denote the
formula $\bigwedge_{C_i \in \mathcal{S}} C_i$.  

Let $\mathcal{S} = \{C_1, \ldots C_r\}$ be a set of clauses. Abusing
notation introduced in Section~\ref{sec:prelim}, let $\atoms{C_i} =
\{z \mid z \in \bX \cup \bY, \lits{C_i}\cap \{z, \neg z\} \neq
\emptyset\}$.  We define an undirected graph $G_{\mathcal{S}} =
(V_{\mathcal{S}}, E_{\mathcal{S}})$, where $V_{\mathcal{S}} = \{C_1,
\ldots C_r\}$ and $(C_i, C_j) \in E_{\mathcal{S}}$ iff $i \neq j$ and
$\atoms{C_i} \cap \atoms{C_j} \cap \bX \neq \emptyset$.  Thus, there
exists an edge $(C_i, C_j)$ iff $C_i$ and $C_j$ share an output atom.
Let $\{\mathcal{S}_1, \ldots \mathcal{S}_q\}$ be the set of maximally
connected components (henceforth called {\MCC}s) of $G_{\mathcal S}$.
It is easy to see that $\cnfform{\mathcal{S}} ~\equiv~
\bigwedge_{k=1}^q \cnfform{\mathcal{S}_k}$; moreover, the output
supports of $\cnfform{\mathcal{S}_k}$ for $k \in \{1, \ldots q\}$ are
mutually disjoint.  We use $C_i \sim_{\mathcal{S}} C_j$ to denote that
clauses $C_i$ and $C_j$ are in the same {\MCC} of $G_{\mathcal{S}}$.
We will soon see how factoring $\cnfform{\mathcal{S}}$ based on
{\MCC}s of $G_{\mathcal S}$ allows us to decompose the
{\CNF}-to-{\syn} compilation problem into independent sub-problems,
thanks to Proposition~\ref{prop:refine-props}(\ref{prop:c})b. Note
that factoring based on {\MCC}s has also been used in
{\dsharp}~\cite{dsharp} for converting a {\CNF} formula to {\dDNNF}.
However, unlike $G_{\mathcal S}$ above, the underlying graph in
{\dsharp} has an edge between every pair of clauses that shares any
atom, including input variables. Thus, $G_{\mathcal S}$ has
potentially fewer edges, and hence smaller {\MCC}s, than the
corresponding graph constructed by {\dsharp}.
\begin{algorithm}[t]
  \scriptsize
  \caption{\FindFDOutsAndRefine}
    \label{alg:fdoutsrefine}
    \KwIn{$\mathcal{S}$: set of clauses,~~$(\bT, {\Def}_{\bT})$: acyclic {\fdefs} in $\cnfform{\mathcal{S}}$}
    \KwOut{$\mathcal{S}'$: set of clauses s.t. $\cnfform{\mathcal{S}'} \preceq_{syn} \cnfform{\mathcal{S}}$, \\
      \qquad\qquad
      \!\!$(\bT', {\Def}_{\bT'})$: Augmented acyclic {\fdefs} in $\cnfform{\mathcal{S}'}$}

    $\mathbf{Out}$ := $\suprt{\cnfform{\mathcal{S}}} ~\cap~ \bX$\;
    $\mathcal{S}'$ := $\mathcal{S}$;~~$(\bT', {\Def}_{\bT'})$ := $(\bT, {\Def}_{\bT})$\tcc*[r]{initialization}
    \Repeat{either $\bT'$ or $\mathcal{S}'$ changes}{
      $(\bT', {\Def}_{\bT'})$ := {\FindFDOuts}($\mathcal{S}',\bT',{\Def}_{\bT'}$)\;
      Let $F$ be the formula $\cnfform{\mathcal{S}'}$\;
      \ForEach{$x_i \in \mathbf{Out}\setminus\bT'$}{
        \uIf{$\theta_{F,\bT',x_i,0}$ is a tautology}{
          $\mathcal{S}'$ := $\mathcal{S}'|_{x_i=1} \cup \{x_i\}$; ~~$\bT' = \bT' \cup \{x_i\}$\;
          ${\Def}_{\bT'}$ := ${\Def}_{\bT'} \wedge (x_i \Leftrightarrow 1)$\;
        }
        \uElseIf{$\theta_{F,\bT',x_i,1}$ is a tautology}{
          $\mathcal{S}'$ := $\mathcal{S}'|_{x_i=0} \cup \{\neg x_i\}$;~~$\bT' = \bT' \cup \{x_i\}$\;
          ${\Def}_{\bT'}$ := ${\Def}_{\bT'} \wedge (x_i \Leftrightarrow 0)$\;
        }
      }
    }   
    \Return $(\mathcal{S}', \bT', {\Def}_{\bT'})$\;
    \normalsize
\end{algorithm}
Before delving into Algorithm {\CToSyn}, we first discuss some
important sub-routines used in the algorithm.  Sub-routine
{\FindFDOutsAndRefine} takes as inputs a set $\mathcal{S}$ of clauses
and a (possibly empty) acyclic system of {\fdefs} $(\bT,
{\Def}_{\bT})$ in $\cnfform{\mathcal{S}}$.  It returns a (possibly
augmented) acyclic system of {\fdefs} $(\bT', {\Def}_{\bT'})$ and a
set of clauses $\mathcal{S}'$ such that $\cnfform{\mathcal{S}'}
\preceq_{syn} \cnfform{\mathcal{S}}$ and $\cnfform{\mathcal{S}'}
\Rightarrow {\Def}_{\bT'}$.  Sub-routine {\FindFDOutsAndRefine} works
by iteratively finding new {\FD} ouptut variables and refining the
specification using
Lemma~\ref{lem:refine-further}(\ref{lem:ref-further-2}) whenever
possible.  In the pseudo-code of {\FindFDOutsAndRefine} (see
Algorithm~\ref{alg:fdoutsrefine}), sub-routine {\FindFDOuts} matches a
pre-defined set of clause-patterns in $\mathcal{S}'$ to identify new
{\FD} output variables not already in $\bT'$.  The patterns currently
matched correspond to {\CNF} encodings of the input-output relation of
common Boolean functions, viz. $\mathsf{and}$, $\mathsf{or}$,
$\mathsf{nand}$, $\mathsf{nor}$, $\mathsf{xor}$, $\mathsf{xnor}$,
$\mathsf{not}$ and $\mathsf{identity}$.  For example, we match the
pattern $(\neg \alpha \vee \beta_1)\wedge(\neg \alpha \vee
\beta_2)\wedge (\neg \beta_1 \vee \neg \beta_2 \vee \alpha)$, where
$\alpha, \beta_1, \beta_2$ are place-holders, to identify the
functional definition $(\alpha \leftrightarrow (\beta_1 \wedge
\beta_2))$.  Each new {\FD} output variable thus identified is added
to $\bT'$ and the corresponding functional definition is added to
${\Def}_{\bT'}$ unless this introduces a cyclic dependency among the
{\fdefs} already in ${\Def}_{\bT'}$.  
Assuming all patterns used by {\FindFDOuts} to determine functional
dependencies are sound, the (possibly augmented) $(\bT',
{\Def}_{\bT'})$ computed by {\FindFDOuts} is a system of acyclic
{\fdefs} in $\cnfform{\mathcal{S}'}$.  In lines $6$-$12$ of
Algorithm~\ref{alg:fdoutsrefine}, we next check if
Lemma~\ref{lem:refine-further}(\ref{lem:ref-further-2}) can be applied
to refine $\cnfform{\mathcal{S}'}$ by pivoting on some variable
$x_i \in \mathbf{Out}\setminus\bT'$.
The refinement, if applicable, is easily done by replacing each clause
$C_i \in \mathcal{S}'$ by $C_i|_{x_i=1}$ (resp. $C_i|_{x_i=0}$) and by
adding the unit clause $x_i$ (resp. $\neg x_i$) to $\mathcal{S}'$.
The pivot $x_i$ is also added to $\bT'$ and the corresponding
functional definition ($x_i \Leftrightarrow 1$ or $x_i \Leftrightarrow
0$ as the case may be) is added to ${\Def}_{\bT'}$.

In general, identifying an acyclic system of {\fdefs} in $F$
potentially enables refinement of $F$ via
Lemma~\ref{lem:refine-further}(\ref{lem:ref-further-2}), which in turn, can lead to augmenting
the acyclic system of {\fdefs} further.
%
Therefore, the loop in lines $3$-$13$ of
Algorithm~\ref{alg:fdoutsrefine} is iterated until no new {\FD}
outputs or additional refinements are obtained. Once this happens,
subroutine {\FindFDOutsAndRefine} returns the resulting acyclic system
of {\fdefs} $(\bT', {\Def}_{\bT'})$ and the resulting set of refined
clauses $\mathcal{S}'$.

Two other important sub-routines used in {\CToSyn} are {\GetCkt} and
{\GetCktWithDefs}.  Sub-routine {\GetCkt} takes as input an {\NNF}
formula $G(\bX,\bY)$ and returns the DAG representation of
$G(\bX,\bY)$.  Sub-routine {\GetCktWithDefs} takes as input a system
of acyclic {\fdefs} $(\bT,{\Def}_{\bT})$, where
$\bX \cap \suprt{{\Def}_{\bT}} = \bT$ (i.e.  $\bT$ is the entire
output support of ${\Def}_{\bT}$).  It returns a DAG
representation of a {\syn} specification equivalent to ${\Def}_{\bT}$.
Without loss of generality, let $x_1 \sqsubset \ldots \sqsubset x_n$
be a linear ordering of the output variables in $\bT$ such that the
functional definition of $x_{i}$ in ${\Def}_{\bT}$ does not depend on
any $x_{j}$ for $j \ge i$.  Such an ordering always exists since
$(\bT, {\Def}_{\bT})$ is an acyclic system of {\fdefs}.  Let $x_{i}
\Leftrightarrow \mathsf{op}_{i}(u_1,\ldots u_{n_i})$ be the functional
definition of $x_i$ in ${\Def}_{\bT}$, where $\mathsf{op}_i$ is a
Boolean function identified via clause-pattern matching in sub-routine
{\FindFDOuts}.  For each $i$ in $\sqsubset$-order in $\{1, \ldots
n\}$, we now construct a DAG ${\Ckt}_{i}$ representing
$\mathsf{op}_i(u_1,\ldots u_{n_i})$ in {\NNF}. While constructing
${\Ckt}_{i}$, we ensure that every $x_{j} \in \bT$ that is also an
argument of $\mathsf{op}_i$ is replaced by the root, say $t_j$, of
the DAG ${\Ckt}_{j}$.  Since $(\bT, {\Def}_{\bT})$ is an acyclic
system of {\fdefs}, this is always possible. Finally, we construct the
overall DAG, say {\Ckt}, representing
$\bigwedge_{x_i \in \bT}\left((x_i \wedge t_i) \vee (\neg x_i \wedge
\neg t_i)\right)$.   It is
easy to see that for every $x_i \in \bT$, there are no paths from
$x_i$ and $\neg x_i$ that meet for the first time at an $\wedge$-labeled node
in {\Ckt}.  Abusing notation and using {\Ckt} to denote the
specification represented by the above DAG, we therefore have
$\decomp{i}{\Ckt}$ is $\wedge_i$-unrealizable for all $i \in \{1,
\ldots n\}$; hence ${\Ckt}$ is in {\syn}.  


We are now in a position to describe Algorithm {\CToSyn}.  The
algorithm is recursive and takes as inputs a set $\mathcal{S}$ of
clauses, a (possibly empty) system of acyclic {\fdefs} $(\bT,
\Def_{\bT})$ in $\cnfform{\mathcal{S}}$, and the recursion level
$\ell$.  Initially, {\CToSyn} is invoked with $\mathcal{S} =$
given set of {\CNF} clauses, $\bT = \emptyset$, ${\Def}_{\bT} = 1$
and $\ell=0$.  The pseudocode of {\CToSyn}, shown in
Algorithm~\ref{alg:c2syn}, first computes the output support
$\mathbf{Out}$ of $\cnfform{\mathsf{S}}$, and then checks a few
degenerate cases (lines $2$-$8$) to determine if a refined {\syn}
specification can be easily obtained.  
In case these checks fail, sub-routine {\FindFDOutsAndRefine} is invoked to
augment the set $\bT'$ of functionally dependent outputs and their
corresponding acyclic {\fdefs} ${\Def}_{\bT'}$, and also to obtain a
(possibly) refined set $\mathcal{S}'$ of clauses.
If all outputs in $\mathbf{Out}$ get functionally determined by this,
Lemma~\ref{lem:refine-further}(\ref{lem:ref-further-1}) guarantees that
${\Def}_{\mathbf{Out}} \preceq_{syn} \cnfform{\mathcal{S'}}$; hence an
invocation of {\GetCktWithDefs}($\mathbf{Out}, {\Def}_{\mathbf{Out}}$)
gives the desired result in line $12$.
%
%
Otherwise, we check in lines $14$-$17$ if
Theorem~\ref{thm:characmain}(\ref{thm:skfunchar}) can be applied.
Recall that Theorem~\ref{thm:characmain}(\ref{thm:skfunchar}) relaxes
the requirements of the {\syn} definition by requiring
$\wedge_i$-unrealizability only when {\GASCK} functions are
substituted for the $\bX$ variables.  As discussed in
Section~\ref{sec:almost-char}, the relaxed requirement can be checked
by testing the unsatisfiability of the error formula $\varepsilon$ for
the {\GASCK} function vector $\Psi$.  If $\varepsilon$ is indeed
unsatisfiable, $\Psi$ is a Skolem function vector for $\mathbf{Out}$
in $\cnfform{\mathcal{S}'}$, and hence $\bigwedge_{x_i \in \mathbf{Out}}(x_i \Leftrightarrow \Psi_i)$ refines $\cnfform{\mathcal{S}'}$.

If $\varepsilon$ is satisfiable, we use a sub-routine
{\OutVarToBranch} that heuristically chooses an output variable $x \in
\mathbf{Out} \setminus \bT'$ on which to branch.  Currently, we use a
       {\VSIDS}~\cite{vsids} score based heuristic, similar to that
       used in {\dsharp}~\cite{dsharp}, to rank variables in
       $\mathbf{Out}\setminus\bT'$, and then choose the variable with
       the highest score.  This allows us to represent
       $\cnfform{\mathcal{S}'}$ as
       $x_i\wedge\cnfform{\mathcal{S}'|_{x=1}} \vee \neg
       x_i\wedge\cnfform{\mathcal{S}'|_{x = 0}}$, so that we can
       refine the two disjuncts independently, thanks to
       Proposition~\ref{prop:refine-props}(\ref{prop:c})a.  However, this
       may lead to some duplicate processing of clauses.  We can avoid this
       by factoring out the subset of clauses whose satisfiability is
       independent of whether $x_i$ is set to $1$ or $0$.  Let
       $\mathcal{S}_1$ (resp. $\mathcal{S}_2$) be the subset of
       clauses in $\mathcal{S}'$ that are in the same {\MCC} of
       $G_{\mathcal{S}'}$ as some $C_j$ that has $x$ (resp.  $\neg x$)
       as a literal.  Let $\mathcal{S}_3$ be the set of all clauses in
       $\mathcal{S}'$ that are neither in $\mathcal{S}_1$ nor
       $\mathcal{S}_2$.  By definition of $G_{\mathcal{S}'}$, the
       sub-specifications $\cnfform{\mathcal{S}_1}$ and
       $\cnfform{\mathcal{S}_3}$ (and similarly,
       $\cnfform{\mathcal{S}_2}$ and $\cnfform{\mathcal{S}_3}$) do not
       share any output variable in their supports, and can be refined
       independently.  This is exactly what algorthm {\CToSyn} does in
       lines $19$-$30$.  The roots of the DAGs resulting from
       the recursive calls in lines $27$, $28$ and $29$ are finally
       combined as in line $30$ to yield the desired DAG representation.

\begin{algorithm}[t]
  \scriptsize
  \caption{\CToSyn}
  \label{alg:c2syn}
  \KwIn{$\mathcal{S}$: set of clauses,~~$(\bT, {\Def}_{\bT})$: acyclic {\fdefs} in $\cnfform{\mathcal{S}}$,~~$\ell$: recursion level}
  \KwOut{DAG representation of $\uF$ in {\syn} s.t. $\uF \preceq_{syn} \cnfform{\mathcal{S}}$}

  $\mathbf{Out}$ := $\suprt{\cnfform{\mathcal{S}}} \cap \bX$\;
  \If{$\cnfform{\mathcal{S}}$ is valid (resp. inconsistent)}{
    \Return {\GetCkt}($1$) (resp. {\GetCkt}($0$))\;
  }
  \ElseIf{$\cnfform{\mathcal{S}}$ is semantically independent of inputs $\bY$}{
      Let $\pi$ be a satisfying assignment of $\cnfform{\mathcal{S}}$\;
      \Return {\GetCkt}($\form{\proj{\pi}{\mathbf{Out}}}$)\;
  }
  \ElseIf{$\cnfform{\mathcal{S}}$ is semantically independent of $\mathbf{Out}$}{
    \Return {\GetCkt}($1$)\;
  }
  \Else{
    $(\bT',{\Def}_{\bT'}, \mathcal{S}')$ :=
    {\FindFDOutsAndRefine}($\mathcal{S}, \bT, {\Def}_{\bT}$)\;

    \If{$\mathbf{Out}\setminus\bT' = \emptyset$}{
      \Return {\GetCktWithDefs}($\mathbf{Out}, {\Def}_{\mathbf{Out}}$)\;
    }
    \Else{
      Let $\Psi$ be {\GASCK} Skolem function vector for $\mathbf{Out}$ in $\cnfform{\mathcal{S}'}$\;
      Let $\varepsilon$ := $\cnfform{\mathcal{S}'}(\mathbf{Out},\bY) \wedge \neg\cnfform{\mathcal{S}'}(\mathbf{Out}',\bY) \wedge \bigwedge_{x_i \in \mathbf{Out}}\left(x_i' \Leftrightarrow \Psi_i\right)$\tcc*[r]{error formula for $\Psi$, as in~\cite{fmcad2015:skolem}}

      \If{$\varepsilon$ is unsat} {
        \Return {\GetCktWithDefs}($\mathbf{Out}, \bigwedge_{x_i \in \mathbf{Out}}(x_i \Leftrightarrow \Psi_i)$)\;
      }
      
      $x$ := {\OutVarToBranch}($\mathcal{S}', \mathbf{Out}\setminus{\bT'}$)\;
      Pos := $\{C_j \in \mathcal{S}' \mid x \in \lits{C_j}\}$\;
      Neg := $\{C_j \in \mathcal{S}' \mid \neg x \in \lits{C_j}\}$\;
      $\mathcal{S}_1$ := $\{C_i \in \mathcal{S}' \mid \exists C_j \in \mathrm{Pos}\, \left(C_i \sim_{\mathcal{S}'} C_j\right) \}$\;
      $\bT_1$ := $\bT' ~\cap~ \suprt{\cnfform{\mathcal{S}_1}}$\;
      $\mathcal{S}_2$ := $\{C_i \in \mathcal{S}' \mid \exists C_j \in \mathrm{Neg}\, \left(C_i \sim_{\mathcal{S}'} C_j\right) \}$\;
      $\bT_2$ := $\bT' ~\cap~ \suprt{\cnfform{\mathcal{S}_2}}$\;
      $\mathcal{S}_3$ := $\{C_i \in \mathcal{S}' \mid \forall C_j \in \mathrm{Pos} \cup \mathrm{Neg}\, \left(C_i \not\sim_{\mathcal{S}'} C_j\right)$\;
      $\bT_3$ := $\bT' ~\cap~ \suprt{\cnfform{\mathcal{S}_3}}$\;

    Let $t_1$ := root of
    {\CToSyn}$(\mathcal{S}_1|_{x=0}, \bT_1, {\Def}_{\bT_1}|_{x=0}, \ell+1)$\;
    Let $t_2$ := root of
    {\CToSyn}$(\mathcal{S}_2|_{x=1}, \bT_2, {\Def}_{\bT_2}|_{x=1}, \ell+1)$\;
    Let $t_3$ := root of
    {\CToSyn}$(\mathcal{S}_3, \bT_3, {\Def}_{\bT_3}, \ell+1)$\;
    \Return {\GetCkt}($t_3 \wedge ((x \wedge t_2) \vee (\neg x \wedge t_1))$)
    }
  }
  \normalsize
\end{algorithm}

\begin{restatable}{theorem}{ctosyncorrect}
  For every set $\mathcal{S}$ of clauses, {\CToSyn}$\left(\mathcal{S},
  \emptyset, 1, 0\right)$ always terminates and returns a DAG
  representation of a {\syn} specification $\uF$ such that $\uF
  \preceq_{syn} \cnfform{\mathcal{S}}$.
\end{restatable}

\section{Experimental results} 
\label{sec:expt}
We ran Algorithm {\CToSyn} on a suite of {\CNF} specifications
comprised of benchmarks from the Prenex 2QBF track of $\qbf$
2018~\cite{qbfeval2018}, and the {\tt .qdimacs} version of $\fact$
benchmarks~\cite{cav18}, which we will refer to as $\factqd$. By Theorem~\ref{thm:succinctness}(\ref{thm:succinct-other-dds}),
a {\BDD} specification can be compiled to an
equivalent {\syn} specification in linear time.  Therefore, any
algorithm that compiles a {\CNF} specification to an ROBDD can be
viewed as an alternative to {\CToSyn} for compiling a {\CNF}
specification to {\syn} (albeit without refinement).  We compare the
performance of {\CToSyn} with that of a BDD compiler and two
state-of-the-art boolean function synthesis tools, namely, $(i)$ the
AIG-NNF pipeline of $\obfss$~\cite{cav18} with ABC's MiniSat as the
SAT solver and $(ii)$ $\cadet$~\cite{Rabe16,RTRS18}.  For the BDD Compiler,
the {\tt .qdimacs} input was converted to an AIG using simple Tseitin
variable detection; this AIG was then simplified and ROBDDs built
using dynamic variable ordering (of all input and output variables) --
this is part of the BDD pipeline of $\obfss$~\cite{cav18}, henceforth
called $\bddbfss$. We also ran {\dsharp}~\cite{dsharp} which compiles
a {\CNF} formula into \dDNNF{} (and hence {\syn} by
Theorem~\ref{thm:succinctness}(\ref{thm:succinct-other-dds})), but it
was successful on very few of our benchmarks; hence we do not present
its performance. Each tool took as input the same {\tt .qdimacs} file.  Experiments
were performed on a cluster with $20$ cores and $64$ GB memory per
node, each core being a $2.2$ GHz Intel Xeon processor running
CentOS6.5. Each run was performed on a single core, with timeout of
$1$ hour and main memory limited to $16$GB.

For {\CToSyn}, several benchmarks were solved in the initial part of the Algorithm~\ref{alg:c2syn} before line 17, i.e., before any recursive calls are made.  Table~\ref{tab:c2syn} presents the results for {\CToSyn}, divided into those that succeeded at recursion level zero (Stage-I) and those that required recursions (Stage-II), as well as the comparison with $\bddbfss$. Since BDDs are also in $\syn$, the total number of benchmarks in $\qbf$ which could be compiled into $\syn$ (by either compiler) is a whopping $283/402$.
\begin{table}[htb]
    \centering
    \scalebox{0.8} {
    \begin{tabular}{|c|c|c|c|c|c|}
    \hline
        {Benchmarks}&
        \multicolumn{3}{c|}{Compiled By \CToSyn} & {BDD} & {Total} \\
        \cline{2-4}
        {(Total)} & 
        Stage I & Stage II & Total &  {compilation} & {in \syn} \\

  \hline
       $\qbf$ (402) & 103 & 82 & 185 &  153 & \bf{283}   \\
   \hline
       $\factqd$ (6) & 0 &  6 & 6 & 6 & \bf{6}   \\
       \hline
    \end{tabular}
    }
    \caption{Compilation into $\syn$}
    \label{tab:c2syn}
\end{table}

\begin{figure}
\centering
  \includegraphics[angle=-90,scale=0.3] {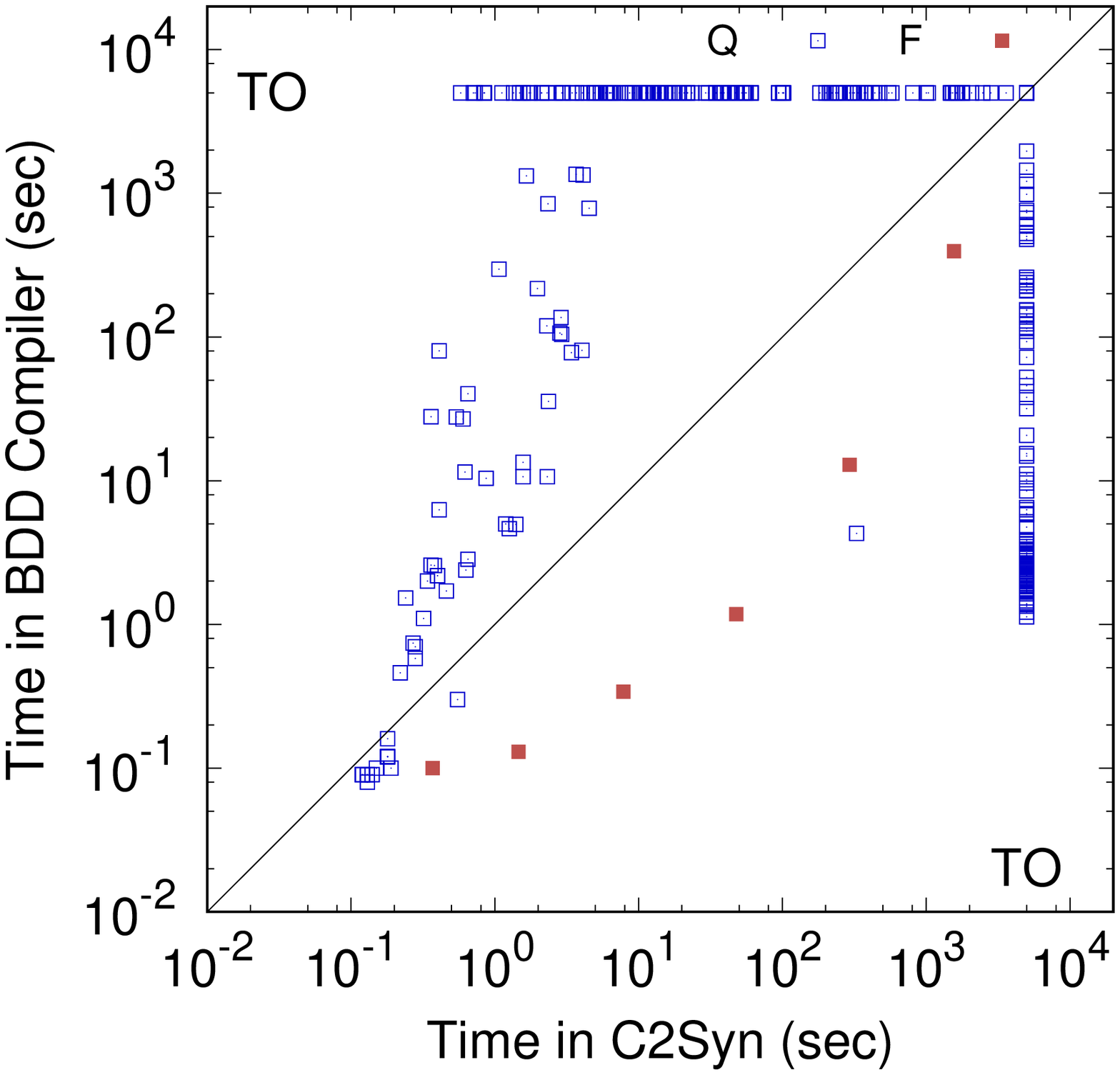} 
  \caption{Performance of {\CToSyn} and $\bddbfss$}
\label{c2syn:bdd}
\end{figure}

Figure~\ref{c2syn:bdd}
compares the run-times of {\CToSyn} and $\bddbfss$:
for most $\qbf$ benchmarks that were solved by both, {\CToSyn} took less time,
while for $\factqd$, {\CToSyn} took more time.
There were $130$ $\qbf$ benchmarks that {\CToSyn} solved by
$\bddbfss$ couldn't, whereas $98$ were solved by $\bddbfss$ but not {\CToSyn}. This indicates that the two approaches to {\syn} compilation have orthogonal strengths.

\begin{table}[htb]
    \centering
    \scalebox{0.9}
    {
    \begin{tabular}{|c|c|c|c|c|c|}
    \hline
        \multirow{3}{*}{Bench} & \multicolumn{2}{|c|}{{\CToSyn} vs $\cadet$} & \multicolumn{2}{c|} {{\CToSyn} vs {\bfss}} & {{\CToSyn} $\setminus$} \\
        \cline{2-5}
        {} & {{\CToSyn}$\setminus$} & {$\cadet \setminus$} & {{\CToSyn}$\setminus$}   & {$\obfss \setminus$ } & { ($\cadet~ \cup$}  \\
      {mark} & $\cadet$ & {\CToSyn}&  $\obfss$  & {\CToSyn} & {$\obfss$)} \\ 

  \hline
       $\qbf$ &   77 & 105  & 83 & 78 & \bf{74}\\
       $\factqd$ &   2 & 0 & 3 & 0 & \bf{2} \\
       \hline
    \end{tabular}
    }
    \caption{Comparison Results of {\CToSyn}}
    \label{tab:comp}
\end{table}

\begin{figure}
\scalebox{0.6} {
 \hspace*{-1.5cm}
\begin{minipage}[t]{0.4\textwidth}
\begin{subfigure}{1.2in}
  \includegraphics[angle=-90,scale=0.4] {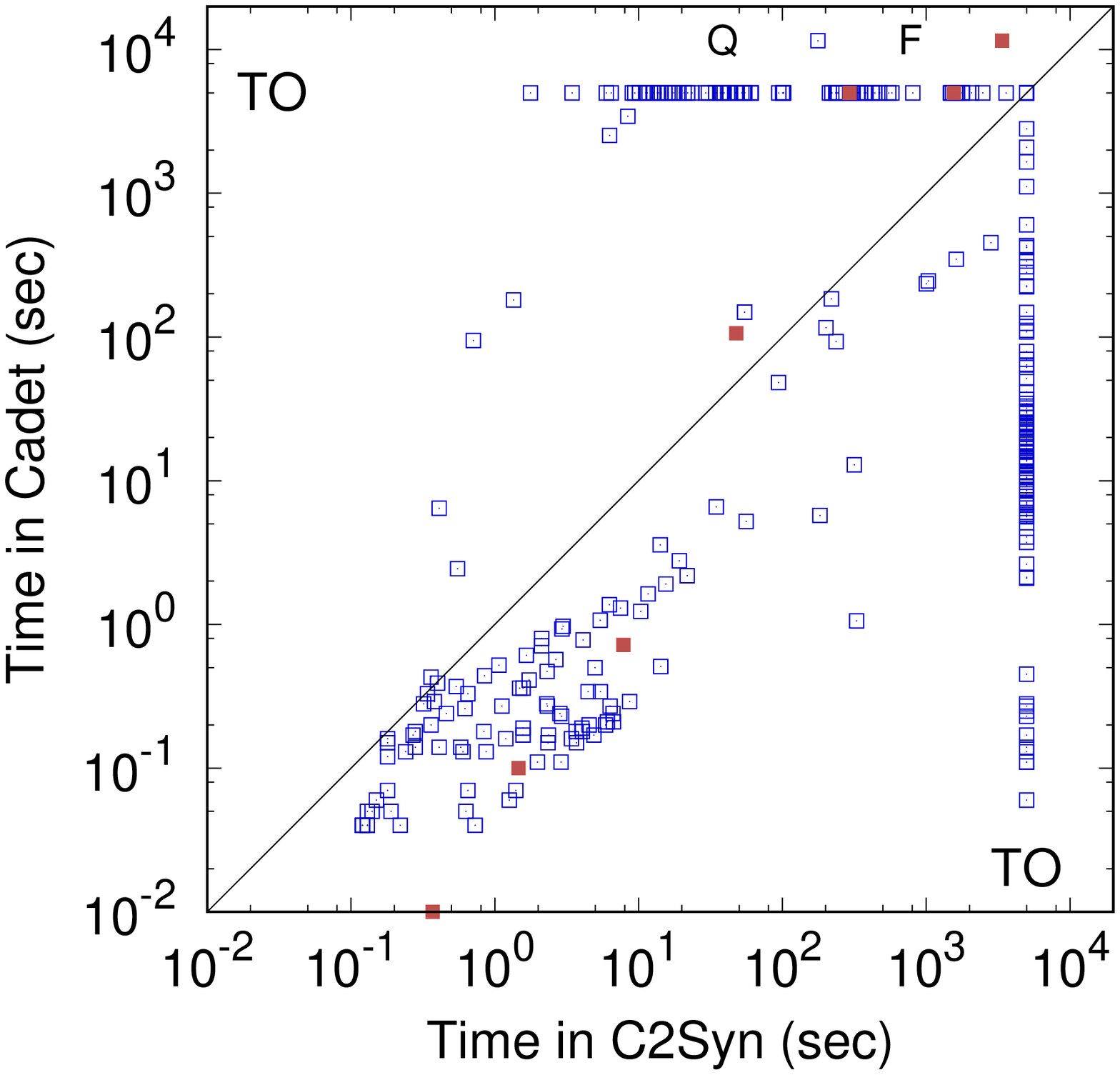} 
\label{fig:cadet}
\end{subfigure}
\end{minipage}
\begin{minipage}[t]{0.3\textwidth}
\begin{subfigure}{1in}
  \includegraphics[angle=-90,scale=0.4] {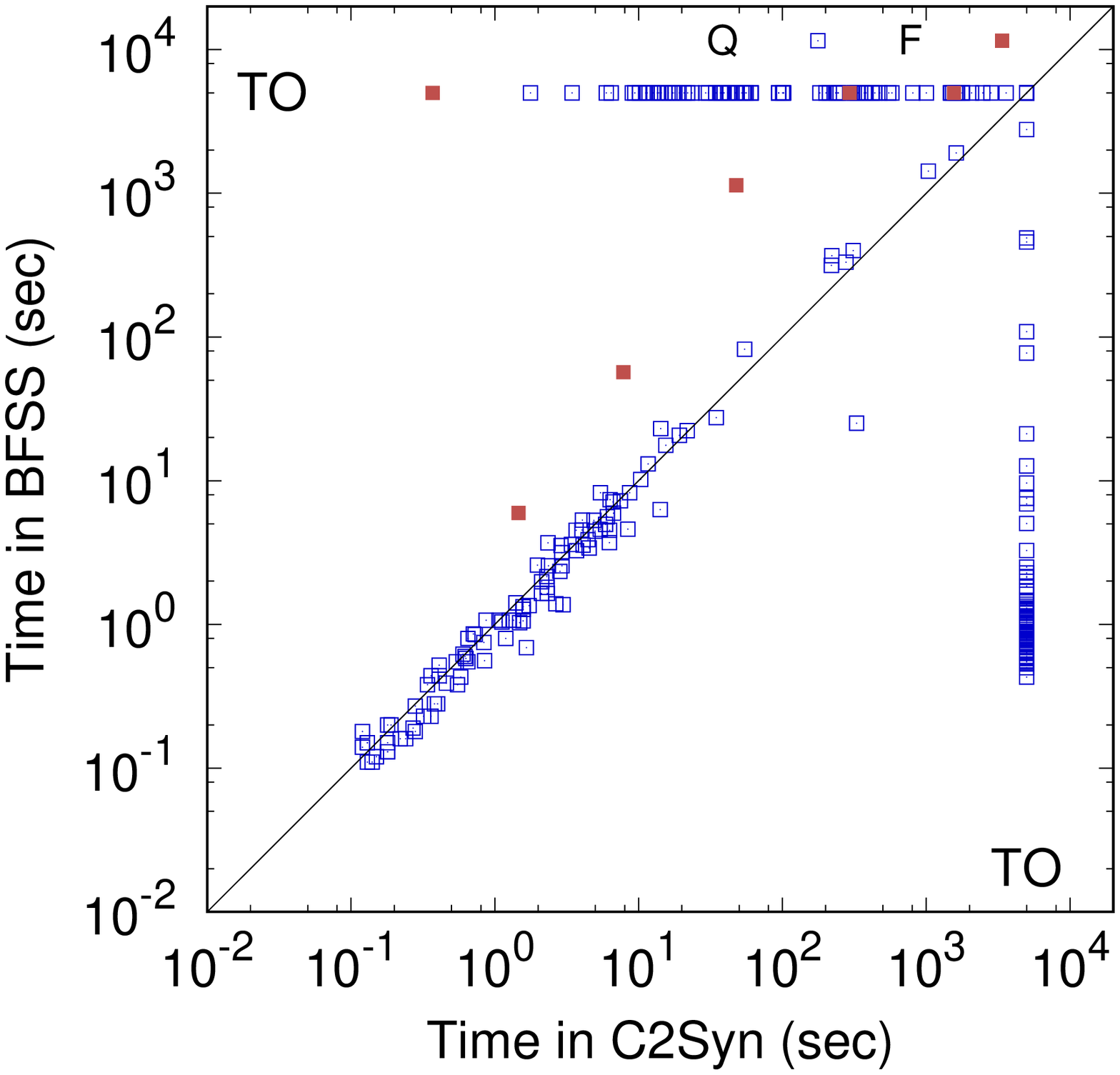} 
\label{fig:bfss}
\end{subfigure}
\end{minipage}
}
\caption{Time comparisons: {\CToSyn} vs $\cadet$ (left) and $\obfss$ (right)}
\label{fig:bfsscadet}
\end{figure}
We next compare {\CToSyn} with $\cadet$ and $\obfss$. $\cadet$ (resp. $\bfss$) solved $213$ (resp. $181$) benchmarks in $\qbf$ and $4$ (resp. $3$) in $\factqd$. Table \ref{tab:comp} gives a comparison in terms of number of benchmarks solved by each tool but not by others. Figure \ref{fig:bfsscadet} (left, right) compares the run-times of {\CToSyn} and
those of $\cadet$ and $\bfss$, respectively.  As expected, since {\CToSyn} does
complete compilation, it takes more time than $\cadet$ and marginally
more than $\bfss$ on many benchmarks, though for most of these, the time taken is less than a minute.  In fact for $\factqd$,
{\CToSyn} takes less time than $\bfss$ on all benchmarks.  Overall,
{\CToSyn} appears to have strengths orthogonal to $\bddbfss$, $\bfss$
and $\cadet$, and adds to the repertoire of state-of-the-art
tools for Boolean functional synthesis.

To validate our experimental results, we also developed an independent approach to verify if the output of {\CToSyn} is (i) in {\syn} and (ii) a refinement of the original specification (which by Theorem 1 and Lemma 4 suffices to efficiently generate Skolem functions). For (i), we check a stronger than required, syntactic condition for being in {\syn}, namely, for every output variable $x_i$, there is no pair of paths from $x_i$ and $\overline{x_i}$ in the DAG output by {\CToSyn} that meet at an $\wedge$-node. Note that this is the sufficient that was described just after Definition~\ref{def:alphadef} in Section~\ref{sec:compiled}. While this requirement is stronger than the semantic requirement for {\syn}, we choose to use this because of the efficient manner in which this can be checked.

For (ii), we just check the two semantic conditions in Definition~\ref{def:refine}. Checking condition (a) requires the use of a 2QBF solver, while condition (b) can be checked using a propositional (un)satisfiability solver. Of the 185 benchmarks on which C2Syn was successful, our verifier successfully verified 183 benchmarks, ran out of memory on 1 and out of time on another benchmark (time limit: 1 hour, main memory limit : 16GB). 

Finally, we note that pre-processing techniques are known to effectively simplify several QBF problem instances. Stage-I of {\CToSyn} can be seen as subsuming several simple QBF preprocessing techniques, e.g., unit clause and pure literal detection, semantic unateness and identifying Tseitin variables. Using more aggressive QBF preprocessing could further improve the performance of our tool, and we leave this for future work.

\section{Conclusion}
\label{sec:concl}
We presented a new sub-class of {\NNF} called {\syn} that admits
quadratic-time synthesis and linear-time existential quantification of
a set of variables.  Our prototype compiler is able to handle several
benchmarks that cannot be handled by other state-of-the-art tools.
Since representations like {\ROBDD}s, {\DNNF} and the like are either
already in or easily transformable to {\SynNNF}, our work is widely
applicable and can be used in tandem with other techniques. As future
work, we intend to work on optimizing our {\syn} compiler.

\bibliographystyle{plain}
\bibliography{ref}

\begin{thebibliography}{10}

\bibitem{cav18}
S.~Akshay, Supratik Chakraborty, Shubham Goel, Sumith Kulal, and Shetal Shah.
\newblock {What's Hard About Boolean Functional Synthesis?}
\newblock In {\em Computer Aided Verification - 30th International Conference,
  {CAV} 2018, Held as Part of the Federated Logic Conference, FloC 2018,
  Oxford, UK, July 14-17, 2018, Proceedings, Part {I}}, pages 251--269, 2018.

\bibitem{tacas2017}
S.~Akshay, Supratik Chakraborty, Ajith~K. John, and Shetal Shah.
\newblock {Towards Parallel Boolean Functional Synthesis}.
\newblock In {\em {TACAS} 2017 Proceedings, Part {I}}, pages 337--353, 2017.

\bibitem{boole1847}
G.~Boole.
\newblock {\em The Mathematical Analysis of Logic}.
\newblock Philosophical Library, 1847.

\bibitem{bryant1986}
R.~E. Bryant.
\newblock Graph-based algorithms for boolean function manipulation.
\newblock {\em IEEE Trans. Comput.}, 35(8):677--691, August 1986.

\bibitem{bryant91}
Randal~E. Bryant.
\newblock On the complexity of {VLSI} implementations and graph representations
  of boolean functions with application to integer multiplication.
\newblock {\em {IEEE} Trans. Computers}, 40(2):205--213, 1991.

\bibitem{CD97}
Marco Cadoli and Francesco~M. Donini.
\newblock A survey on knowledge compilation.
\newblock {\em {AI} Commun.}, 10(3-4):137--150, 1997.

\bibitem{CFTV18}
Supratik Chakraborty, Dror Fried, Lucas~M. Tabajara, and Moshe~Y. Vardi.
\newblock Functional synthesis via input-output separation.
\newblock In {\em 2018 Formal Methods in Computer Aided Design, {FMCAD} 2018,
  Austin, TX, USA, October 30 - November 2, 2018}, pages 1--9, 2018.

\bibitem{bafsyn:fmcad2018}
Supratik Chakraborty, Dror Fried, Lucas~M. Tabajara, and Moshe~Y. Vardi.
\newblock Functional synthesis via input-output separation.
\newblock In {\em 2018 Formal Methods in Computer Aided Design, {FMCAD} 2018,
  Austin, TX, USA, October 30 - November 2, 2018}, pages 1--9, 2018.

\bibitem{darwiche-jacm}
Adnan Darwiche.
\newblock Decomposable negation normal form.
\newblock {\em J. {ACM}}, 48(4):608--647, 2001.

\bibitem{DM11}
Adnan Darwiche and Pierre Marquis.
\newblock A knowledge compilation map.
\newblock {\em CoRR}, abs/1106.1819, 2011.

\bibitem{rsynth}
Dror Fried, Lucas~M. Tabajara, and Moshe~Y. Vardi.
\newblock {BDD}-based boolean functional synthesis.
\newblock In {\em Computer Aided Verification - 28th International Conference,
  {CAV} 2016, Toronto, ON, Canada, July 17-23, 2016, Proceedings, Part {II}},
  pages 402--421, 2016.

\bibitem{Jian}
J.-H.~R. Jiang.
\newblock Quantifier elimination via functional composition.
\newblock In {\em Proc. of CAV}, pages 383--397. Springer, 2009.

\bibitem{jiang2}
J.-H.~R. Jiang and V~Balabanov.
\newblock {Resolution proofs and {S}kolem functions in {QBF} evaluation and
  applications}.
\newblock In {\em Proc. of CAV}, pages 149--164. Springer, 2011.

\bibitem{fmcad2015:skolem}
A.~John, S.~Shah, S.~Chakraborty, A.~Trivedi, and S.~Akshay.
\newblock Skolem functions for factored formulas.
\newblock In {\em FMCAD}, pages 73--80, 2015.

\bibitem{KMPS10}
V.~Kuncak, M.~Mayer, R.~Piskac, and P.~Suter.
\newblock Complete functional synthesis.
\newblock {\em SIGPLAN Not.}, 45(6):316--329, June 2010.

\bibitem{LLM11}
J{\'{e}}r{\^{o}}me Lang, Paolo Liberatore, and Pierre Marquis.
\newblock Propositional independence - formula-variable independence and
  forgetting.
\newblock {\em CoRR}, abs/1106.4578, 2011.

\bibitem{lowenheim1910}
L.~Lowenheim.
\newblock {{\"U}ber} die {Aufl{\"o}sung} von {Gleichungen} in {Logischen}
  {Gebietkalkul}.
\newblock {\em Math. Ann.}, 68:169--207, 1910.

\bibitem{bierre}
Martina~Seidl Marijn~Heule and Armin Biere.
\newblock {Efficient Extraction of Skolem Functions from QRAT Proofs}.
\newblock In {\em Formal Methods in Computer-Aided Design, {FMCAD} 2014,
  Lausanne, Switzerland, October 21-24, 2014}, pages 107--114, 2014.

\bibitem{vsids}
Matthew~W. Moskewicz, Conor~F. Madigan, Ying Zhao, Lintao Zhang, and Sharad
  Malik.
\newblock Chaff: Engineering an efficient sat solver.
\newblock In {\em Proceedings of the 38th Annual Design Automation Conference},
  DAC '01, pages 530--535, New York, NY, USA, 2001. ACM.

\bibitem{dsharp}
Christian Muise, Sheila~A. McIlraith, J.~Christopher Beck, and Eric Hsu.
\newblock {DSHARP: Fast d-DNNF Compilation with sharpSAT }.
\newblock In {\em AAAI-16 Workshop on Beyond NP}, 2016.

\bibitem{qbfeval2018}
QBFLib.
\newblock {QBFE}val 2018.
\newblock {http://www.qbflib.org/qbfeval18.php}.

\bibitem{Rabe16}
M.~N. Rabe and S.~A. Seshia.
\newblock Incremental determinization.
\newblock In {\em Theory and Applications of Satisfiability Testing - {SAT}
  2016 - 19th International Conference, Bordeaux, France, July 5-8, 2016,
  Proceedings}, pages 375--392, 2016.

\bibitem{Rabe15}
M.~N. Rabe and L.~Tentrup.
\newblock {CAQE:} {A} certifying {QBF} solver.
\newblock In {\em Formal Methods in Computer-Aided Design, {FMCAD} 2015,
  Austin, Texas, USA, September 27-30, 2015.}, pages 136--143, 2015.

\bibitem{RTRS18}
Markus~N. Rabe, Leander Tentrup, Cameron Rasmussen, and Sanjit~A. Seshia.
\newblock Understanding and extending incremental determinization for 2qbf.
\newblock In {\em Computer Aided Verification - 30th International Conference,
  {CAV} 2018, Held as Part of the Federated Logic Conference, FloC 2018,
  Oxford, UK, July 14-17, 2018, Proceedings, Part {II}}, pages 256--274, 2018.

\bibitem{rsynth:fmcad2017}
Lucas~M. Tabajara and Moshe~Y. Vardi.
\newblock Factored boolean functional synthesis.
\newblock In {\em 2017 Formal Methods in Computer Aided Design, {FMCAD} 2017,
  Vienna, Austria, October 2-6, 2017}, pages 124--131, 2017.

\bibitem{tseitin68}
G.~S. Tseitin.
\newblock On the complexity of derivation in propositional calculus.
\newblock {\em Structures in Constructive Mathematics and Mathematical Logic,
  Part II, Seminars in Mathematics}, pages 115--125, 1968.

\bibitem{Valiant79}
L.~G. Valiant.
\newblock Completeness classes in algebra.
\newblock In {\em Proceedings of the Eleventh Annual ACM Symposium on Theory of
  Computing}, STOC '79, pages 249--261, New York, NY, USA, 1979. ACM.

\end{thebibliography}
\newpage

\onecolumn
\appendix

We present the material in this Appendix in single-column format since
a large number of equations needed in the proof of our main results
are presented much better in single-column format.

\section{Material from Section~\ref{sec:compiled}}

\subsection{Proof of Theorem~\ref{thm:succinctness} of Section~\ref{sec:compiled}}
\label{app:family}

This section is dedicated to the proof of Theorem~\ref{thm:succinctness}. We show that \syn{} is a space-efficient DAG-based representation of boolean functions,  when compared with other representations using \fbdd, \DNNF{} and  \dDNNF. 

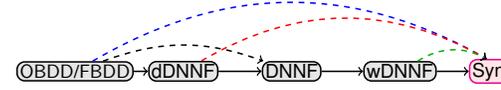
\begin{figure}[t]
\begin{center}
\scalebox{.8}{
\begin{tikzpicture}[->,thick]
\tikzset
  {dest/.style={rectangle,draw=magenta,fill=red!10!white,rounded corners,minimum width=0.5mm,inner sep=0.5mm}}
\tikzset
  {dest1/.style={rectangle,draw=black,fill=black!10!white,rounded corners,minimum width=0.5mm,inner sep=0.5mm}}

\scalebox{0.9}{
 \node[dest1] (s6) at (-6,0) {\obdd/\fbdd};
   \node[dest1] (s5) at (-4,0) {\dDNNF};
   \node[dest1] (s4) at (-2,0) {\DNNF};
   \node[dest1] (s3) at (0,0) {\wDNNF};
  \node[dest] (s) at (2,0) {\syn{}};
  \path(s6) edge[] node[above] {} node{}(s5);
  \path(s5) edge[] node[above] {} node{}(s4);
  \path(s4) edge[] node[above] {} node{}(s3);
  \path(s3) edge[] node[above] {} node{}(s);
  \path(s6) edge[dashed,bend left=20] node[above]{} node{} (s4);
 \path(s6) edge[draw=blue,dashed,bend left=30] node[above]{} node{} (s);
 \path(s5) edge[draw=red,dashed,bend left=30] node[above]{} node{} (s);
 \path(s3) edge[draw=green!70!black,dashed,bend left=30] node[above]{} node{} (s);
}
        \end{tikzpicture}}
\caption{An edge $A$ $\rightarrow$ $B$ means that $A$ is a proper subset of $B$. A blue edge from $A$ to $B$ means $B$ is exponentially more succinct than $A$, while a red edge from $A$ to $B$ means that unless $\pp=\vnp$, $B$ is super-polynomially more succinct than $A$.  The green edge from $A$ to $B$ means that unless $\pp=\np$, $B$ is super-polynomially more succinct than $A$.  The black edge is the exponential succinctness of $\DNNF$ w.r.t \fbdd \cite{darwiche-jacm}.}
\label{fig:nnfzoo}
\end{center}
\vspace*{-6mm}
\end{figure}

First, observe that Part(i) is easy. That is, it has been shown, e.g., in ~\cite{darwiche-jacm} that \fbdd can be converted to \DNNF{} with a linear complexity blowup. Now, focussing on \dDNNF, \DNNF{}, \wDNNF, an examination of their definitions immediately gives us that each of these forms is already in \syn. Further, from the definition again it is clear that \dDNNF is subsumed by \DNNF{}, which is further subsumed by \wDNNF, as depicted in Figure~\ref{fig:nnfzoo}. To show strictness, it suffices to consider Example~\ref{eg:1}, which is in \syn\ but not in \wDNNF\ since $x_2$ and $\neg x_2$ indeed meet up at an $\wedge$-node in $G$. This completes Part (i).

For part (ii), we start by noting that it has been shown in \cite{darwiche-jacm} that the \DNNF{} representation is exponentially more succinct than  \fbdd. We now show that \syn\ is super-polynomially (resp. exponentially) more succinct than \dDNNF, \DNNF{} and \wDNNF{} (resp. \fbdd) representations, unless some long-standing complexity conjectures are falsified. To do this, we describe a family of specifications having a polynomial sized  \syn\ representation, but for which the other representations are necessarily super-polynomially larger, unless these complexity conjectures are falsfied.
 
 Consider the family $\Ff(\bX, \bY)$ of specifications defined as follows. Let $\bX=\{x_1, \dots, x_n\}$, and let     $f_i(\bX_{i+1}^n,\bY)$, $1 \leq i \leq n-1$ be arbitrary boolean functions   in $\NNF$  over $x_{i+1},\dots,x_n, \bY$. 
      We define the family $\Ff(\bX, \bY)_{(\op'_1, \op_1, \dots, \op'_n, \op_n)}$, parametrized 
    by $\op_i \in \{\wedge, \vee\}$, and $\op'_i \in \{\wedge, \vee, \xor\}$ as     
$$(x_1 \op'_1 f_1(\bX_2^n,\bY)) \op_1 
(x_2\op'_2 f_2(\bX_3^n,\bY)) \op_2 \dots \op_{n-1} (x_n \op'_n f_n(\bY)) \op_n f_{n+1}(\bY)$$


\begin{lemma}
Let $g$ be a function in the family of specifications $\Ff(\bX, \bY)_{(\op'_1, \op_1, \dots, \op'_n, \op_n)}$. Then 
\begin{enumerate}
\item 
 If $\op'_1=\dots=\op'_n=\vee$, then $g$ is in  \syn.
\item If $\op'_1=\dots=\op'_n=\xor$, then 	 $g$ is in \syn.
\end{enumerate}
 \label{lem:nf}	
\end{lemma}
\begin{proof}
\begin{enumerate}
\item 	
Let $g$ be any function in the family  with all the $\op'_i=\vee$.
 It is easy to see that $g$ is in \syn, using the sufficient condition in Section~\ref{sec:compiled}. That is, in $\decomp{1}{g}$, there is no $\ol{x_1}$, so we never have a $x_1$ and $\ol{x}_1$ meeting at the root. Further, $\decomp{2}{g}$ after replacing $x_1$ with 1, the leftmost subtree rooted at $\vee$ having children  $x_1, f_1$ is no longer there after constant propagation. In the rest of the tree, we have no occurrences of $\ol{x}_2$, hence no way for $x_2$ and $\ol{x}_2$ to meet at the root. Thus, for each $\decomp{i}{g}$, the argument is similar, since 
on replacing $x_1, \dots, x_{i-1}$ with 1 and doing constant propagation,  the remaining DAG will not have  $x_{i+1}$ and $\ol{x}_{i+1}$ together, which shows that $g$ is in \syn.
\item Let $g$ be any function in the family  with all the $\op'_i=\xor$. Note that in this case, we cannot use the sufficient condition as above, 
clearly, $x_1, \ol{x}_1$ meet at a $\wedge$   in $\decomp{1}{g}$. Nevertheless $g$ is in \syn, if we consider 
  $\decomp{i}{g}$ for $1 \leq i \leq n$, and consider the root node with children $\alpha_1=x_i \vee f_i$ and $\alpha_2=\ol{x}_i \vee \neg f_i$, after 
   substituting $x_1, \dots, x_{i-1},\ol{x}_1, \dots, \ol{x}_{i-1}$ to 1, 
   $\ol{x}_{i+1}, \dots, \ol{x}_n$ to $\neg x_{i+1}, \dots, \neg x_n$, 
    and constant propagation, 
    it is easy to see that $\alpha_1^{11} \wedge \alpha_2^{11}=1 ~ \op_1 G$, 
    $\alpha_1^{10} \wedge \alpha_2^{10}=\neg f_i\op_1 G$ and $\alpha_1^{01} \wedge \alpha_2^{01}=f_i\op_1 G$ for $\op_1\in \{\vee,\wedge\}$ and some $G$. Thus  $((\alpha_1^{11} \wedge \alpha_2^{11}) \wedge \neg (\alpha_1^{10} \wedge \alpha_2^{10}) \wedge \neg(\alpha_1^{01} \wedge \alpha_2^{01}))$ is unsatisfiable. Thus, $g$ is $\wedge_i$ unrealizable for any $i$. 
  \end{enumerate}
 \end{proof}

\begin{theorem}[Restatement of Theorem~\ref{thm:succinctness}(ii)]
\begin{enumerate}
\item[(a)] 	
  There are functions which admit polynomial sized \syn{} representations, yet admit only exponential sized \fbdd{} representations.
\item[(b)] Unless $\pp=\vnp$, there are functions which admit polynomial sized \syn\ representations, yet admit only super-polynomial sized \dDNNF{} representations.
\item[(c)] Unless $\pp=\np$, there are functions which admit polynomial sized \syn\ representations, yet admit only super-polynomial sized \wDNNF{} and \DNNF{} representations. 
\end{enumerate}
\end{theorem}
\begin{proof}
We use the family of specifications $\Ff(\bX, \bY)$ defined above, with different instantiations to obtain all three results.  Set $\op_1=\dots=\op_n=\wedge$, $\op'_1=\dots=\op'_n=\vee$, $f_i(\bX_{i+1}^n,\bY)=\top$ for 
$2 \leq i \leq n$, obtaining $g=x_1 \vee f_1(\bX_{2}^n,\bY)$. 
Let $\bY=\{y_1, \dots, y_{n-1}\}$. As seen in Lemma \ref{lem:nf},
$g$ is in $\syn{}$. In each of the subparts below, we define $f_1(\bX_{2}^n,\bY)$ appropriately.   

\medskip 
\noindent{Item (a): \bf{Succinctness w.r.t \fbdd}}.  Let $k=n-1$. We use the $k$-bit multiplier function 
over $\{x_2, \dots, x_n\} \cup \{y_{1}, \dots, y_{n-1}\}$ in the construction of  $f_1(\bX_{2}^n,\bY)$. 
The two $k$ bit arguments to the multiplier are respectively, 
$\{x_2, \dots, x_n\}$ and $\{y_{1}, \dots, y_{n-1}\}$
with $x_n, y_{n-1}$ being the most significant bits, and 
$x_2, y_{1}$ being the least significant bits. 
  Let $f_1(\bX_{2}^n,\bY)$ be the boolean function representing the 
  $k$th bit of the  $k$-bit multiplier function. 
  The size of $f_1(\bX_{2}^n,\bY)$ is quadratic in $k$, since the size of any multiplier circuit 
consisting of $\vee, \wedge$ gates is quadratic in $k$ (sum of $k^2$ partial products).
For this $f_1(\bX_{2}^n,\bY)$, the  size of  $g$ is $\mathcal{O}(k^2+1)$.  

Let $\rep_1$ be a representation of $g$ using  \fbdd,~ by fixing a certain variable order. 
Set $x_1{=}0$. This assignment 
makes $g=f_1(\bX_{2}^n,\bY)$. 
 Indeed, the \fbdd~ representation obtained as a  restriction of $\rep_1$
with respect to this  truth assignment  is simpler \cite{bryant1986}. It is known \cite{bryant91} that any \fbdd, \obdd~ representations 
for $f_1(\bX_{2}^n,\bY)$ is exponential in $k$. This establishes the exponential succinctness 
of \syn\ over \fbdd. 

\medskip 
\noindent{Item (b): \bf{Succinctness w.r.t \dDNNF}}.  
 We use a CNF encoding of the perfect matchings of a bipartite graph $G$ (denoted $\pma(G)$) 
in the construction of $f_1(\bX_{2}^n,\bY)$. 
Given a bipartite graph $G$ with two parts $U=\{u_1, \dots, u_m\}$ and $V=\{v_1, \dots, v_m\}$, 
we can define a 0-1 matrix $A=(a_{ij}), 1 \leq i, j \leq m$ such that $a_{ij}=1$ iff 
there is an  edge between $u_i \in U$ and $v_j \in V$. 
 It is easy to see from the definition of the permanent of $A$ (denoted $\perm(A)$) that $\perm(A)=\pma(G)$.   
Likewise, the number of perfect matchings of a bipartite graph is the permanent of its incidence matrix.  
Set $f_1(\bX_{2}^n,\bY)$ as  the function which encodes $\pma(G)$. 

Let $\rep_2$ be the \dDNNF~representation   
of $g$. As in the first case, choose an assignment $x_1{=}0$  obtaining $g=0 \vee f_1(\bX_{2}^n,\bY)=f_1(\bX_{2}^n,\bY)$. 
Then it can be seen that the number of solutions of $f_1$ is exactly the number of perfect matchings  of the bipartite graph $G$. Fixing the assignment $x_1{=}0$ results in a simpler \dDNNF~ representation (say $\rep_3$)
for $g$ (now $f_1$). Counting    
the models of $\rep_3$ can be done in 
time polynomial in the size of $\rep_3$ \cite{DM11}. 
This implies that we can 
find the number of perfect matchings 
of the underlying bipartite graph $G$ in time polynomial in the size  
 of $\rep_3$. Unless $\pp=\vnp$, $\rep_3$ cannot have a polynomial representation, 
 since otherwise, we would obtain a polynomial time solution 
 for computing $\perm(A)$.  This shows the super-polynomial succinctness 
 of \syn\ over \dDNNF, unless $\pp=\vnp$.

\medskip

\noindent Item(c): {\bf Succinctness w.r.t \wDNNF\ and \DNNF\ }
Let $\op'_1=\dots=\op'_n=\vee$, $f_i(\bX_{i+1}^n,\bY)=\top$ for $2 \leq i \leq n$, obtaining $g=x_1 \vee f_1(\bX_{2}^n,\bY)$. As shown in Lemma \ref{lem:nf}, $g$ is in \syn, where $f_1(\bX_{2}^n,\bY)$ is an arbitrary SAT formula. If we can obtain a poly-sized \DNNF{} representation for the function $g$, then using the assignment $x_1=0$ in $g$,  we  obtain a $\DNNF$ representation for $f_1(\bX_{2}^n,\bY)$. But it is known \cite{DM11} that consistency checking is poly-time for \DNNF{} representations.  A polynomial sized \DNNF{} representation for $g$  would imply a polynomial  time solution for the satisfiability checking of an arbitrary SAT formula.  Thus, unless $\pp=\np$,  any \DNNF{} representation for $g$ will necessarily be super polynomial. 
   \end{proof}

This completes the proof of Part (ii) of Theorem~\ref{thm:succinctness}.

Part(iii). 
By Theorem~1 of~\cite{cav18}, we know that there exist instances of poly-sized NNF formulas whose Skolem functions are necessarily super-polynomial size (resp. exponential) unless the polynomial hierarchy collapses (resp. the non-uniform exponential hypothesis is falsified). For any such instance, suppose we were able to obtain a poly-sized \syn{} representation, then by Theorem~\ref{thm:synnnf-exists-synth}, we will be able to synthesize polynomial-sized Skolem functions, which contradicts the above.

To see a concrete example where {\syn} is not likely to be succinct, we refer to Theorem 1 of~\cite{cav18}, where a constructive reduction of the parameterized clique problem to {\BFS} was given. The specification, in this case, has a polynomial-sized representation, but unless some long-standing complexity-theory conjectures are violated, it was shown that any Skolem function must have exponential/super-polynomial size. Thus, unless these conjectures are violated, the same specification in {\syn} must also be exponential/super-polynomial sized. 

This proves Part (iii) and completes the proof of this theorem.

Essentially this means that though we obtain succinctness with respect to several known forms (using classical complexity-theoretic results), it is not the case that \syn\ will always be able to produce a poly-sized representation.

\subsection{Proof of Theorem~\ref{thm:characmain}}
\label{app:charac}
Let us recall the characterization theorem from Section~\ref{sec:compiled}.
\characmain*

\begin{proof}
Part 1): The forward direction follows from Theorem~\ref{thm:synnnf-exists-synth}. For the reverse direction, we will prove the contrapositive: if $F$ is not in \syn, i.e., if $\decomp{i}{\Ff}[\bX_{i+1}^n \mapsto \neg \bX_{i+1}^n]$ is not $\wedge_i$-unrealizable  for some $i \in \{1 \ldots n\}$, we will show that for some $i$, $\exists \bX_1^i F(\bX,\bY)\not \Leftrightarrow \decomp{i+1}{F}[\ol{\bX}_{i+1}^n \mapsto \neg\bX_{i+1}^n]$. Fix any $\bY\in \{\bY'\mid \exists \bX', F(\bX',\bY')\}$, i.e., it is a realizable valuation of inputs. Consider $i$ to be the largest index such that $\decomp{i}{F}$ is not $\wedge_i$-unrealizable,
i.e., the corresponding $\zeta$ is satisfiable. As a result, we have $\alpha^{11}=1$, i.e., $\decomp{i+1}{F}[\ol{\bX}_{i+1}^n \mapsto \neg\bX_{i+1}^n]=1$. On the other hand $\alpha^{01}=\nnf{F}(1^{i-1},0,{\bX}_{i+1}^n,1^{i-1},1,\neg {\bX}_{i+1}^n, \bY)=0$ and $\alpha^{10}=\nnf{F}(1^{i-1},1,{\bX}_{i+1}^n,1^{i-1},0,\neg {\bX}_{i+1}^n, \bY)=0$.    By monotonicity, every assignment of $x_1,\ldots x_{i-1},x_i$ will also result in 0 in $\nnf{F}$, which implies that $\exists \bX_1^i F(\bX,\bY)=0$. Thus for this $i$, $\exists \bX_1^i F(\bX,\bY)\not \Leftrightarrow \decomp{i+1}{F}[\ol{\bX}_{i+1}^n \mapsto \neg\bX_{i+1}^n]$, which completes the proof.
Part 2):
{ \bf Forward direction:}  We will prove the contrapositive, i.e., if $\decomp{i}{\Ff}[\bX_{i+1}^n \mapsto \Psi_{i+1}^n]$ is not $\wedge_i$-unrealizable  for some $i \in \{1 \ldots n\}$, we will show that $\Psi_1^n$ is not a correct Skolem function vector for $\bX_1^n$ in $F(\bX, \bY)$. Fix any $\bY\in \{\bY'\mid \exists \bX', F(\bX',\bY')\}$, i.e., it is a realizable valuation of inputs. Consider $i$ to be the largest index such that $\decomp{i}{F}[\bX_{i+1}^n\mapsto {\Psi}_{i+1}^n,\ol{\bX}_{i+1}^n\mapsto \neg {\Psi}_{i+1}^n]$ is not $\wedge_i$-unrealizable, i.e., the corresponding $\zeta$ is satisfiable. 

  We claim that one of the ${\Psi}_{i+1}^n$ must be an incorrect skolem function for this $\bY$. Suppose not, i.e., suppose all of them are correct. Then we have 
  \begin{align}\label{eq:1}
    \exists x_1,\ldots, x_i \nnf{F}(x_1,\ldots,x_i,{\Psi}_{i+1}^n,\neg{x_1},\ldots \neg{x_i},\neg{\Psi}_{i+1}^n,\bY)=1
\end{align}
  However, because at $i$, $\zeta$ is satisfiable, 
   we have $\nnf{F}(1^{i-1},0,{\Psi}_{i+1}^n,1^{i-1},1,\neg {\Psi}_{i+1}^n, Y)=0$ and $\nnf{F}(1^{i-1},1,{\Psi}_{i+1}^n,1^{i-1},0,\neg {\Psi}_{i+1}^n, Y)=0$.
   By monotonicity, every assignment of $x_1,\ldots x_{i-1}$ will also result in 0 in $\nnf{F}$.     But this contradicts~(\ref{eq:1}). Hence all the skolem functions cannot be correct for this $\bY$, proving the forward direction.
  
{\bf Reverse direction:}
Again, we prove by taking the contrapositive. Suppose, ${\Psi}_{i+1}^n$
is not a correct Skolem function vector.  In~\cite{fmcad2015:skolem}, it was shown that for any function vector $\varphi_1^n$, it is a Skolem function vector for $F$ iff the \emph{error formula} $\varepsilon_\varphi \equiv F(\bX,\bY) \wedge \neg F(\bX',\bY) \wedge \bigwedge_{i=1}^n(x_i' \leftrightarrow \varphi_i)$ is unsatisfiable. We will use this characterization now, i.e., since  ${\Psi}_{i+1}^n$ is not a correct Skolem function vector, the error formula $\varepsilon_\Psi$ must be satisfiable.


Hence, we start by considering $\bY^*$ which gives a satisfying assignment for the error formula $\varepsilon_\Psi$. That is, 
  \begin{align}
    \exists \bX' F(\bX',\bY^*)\wedge \exists 1\leq i\leq n ~\neg \exists \bX_{1}^{i-1} F(\bX_1^{i-1},{\Psi}_i^n(\bY^*),\bY^*)\label{eq:pf0}
  \end{align}
  Let $k$ be the highest such $i$ such that the above statement holds. That is, after $k$, the Skolem functions given by $\Psi$ are correct, and at $k$ they are incorrect.   Then, we observe that the value at $k$ must be 1, i.e.,
\begin{align}
  \Psi_k(\bY^*)=\nnf{F}({\mathbf{1}}^{k-1}, 1, {\Psi}_{k+1}^n(\bY^*),{\mathbf{1}}^{k-1}, 0, \neg {\Psi}_{k+1}^n(\bY^*),\bY^*) =1 \label{eq:pf2}
    \end{align} 
To see this, observe that $\exists \bX'F(\bX',\bY^*)$ along with maximality of $k$ implies that $\exists \bX_1^k F(\bX_1^k,{\Psi}_{k+1}^n(\bY^*),\bY^*)=1$, which in turn implies that
\begin{align*}
  \nnf{F}({\mathbf{1}}^{k-1}, 1, {\Psi}_{k+1}^n(\bY^*),{\mathbf{1}}^{k-1}, 0, \neg {\Psi}_{k+1}^n(\bY^*),\bY^*)
  \vee \nnf{F}({\mathbf{1}}^{k-1}, 0, {\Psi}_{k+1}^n(\bY^*),{\mathbf{1}}^{k-1}, 1, \neg {\Psi}_{k+1}^n(\bY^*),\bY^*)=1\end{align*}
  Now, if $\Psi_k(\bY^*)=0$, this implies $\nnf{F}({\mathbf{1}}^{k-1}, 0, {\psi'}_{k+1}^n(\bY^*),{\mathbf{1}}^{k-1}, 1, \neg {\psi'}_{k+1}^n(\bY^*),\bY^*)=1$. But then, setting $x_k=1$ is indeed correct, which would imply that there is no error at $k$, which violates the assumption on $k$. Thus we must have $\Psi_k(\bY^*)=1$.
  
Now, we know that this is an incorrect assignment to $x_k$, which implies that the correct assignment is a $0$ and we know that $ \exists \bX_1^{k-1} F(\bX_1^{k-1},1,{\Psi}_{k+1}^n(\bY^*),\bY^*)$ is a correct assignment to $x_k$. Hence, we must have
\begin{align}
  \exists \bX_1^{k-1} F(\bX_1^{k-1},1,{\Psi}_{k+1}^n(\bY^*),\bY^*)=0\nonumber\\
  \implies \exists \bX_1^{k-1}\nnf{F}(\bX_1^{k-1}, 1,{\Psi}_{k+1}^n(\bY^*),\neg \bX_1^{k-1},0, \neg {\Psi}_{k+1}^n(\bY^*),\bY^*)=0
  \label{eq:pf3}
  \end{align}

The fact that equations~(\ref{eq:pf2}), (\ref{eq:pf3}) hold together imply that the Skolem function $\Psi$ is wrong \emph{at level $k$}, since it gives value 1, but fixing $x_k=1$, there is no way to set lower variables to get 1. The rest of the proof is a careful case-analysis, where we either show that $\zeta$ (with Skolem functions assigned according to $\Psi$) at level $k$ is satisfiable, i.e.,  $\decomp{k+1}{F}[\bX_{k+1}^n\mapsto \Psi_{k+1}^n]$ is not $\wedge_k$-unrealizable and hence the proof terminates, 
or we show that these equations are satisfied at a lower level (i.e., there is an error at a lower level). Since number of levels is finite this procedure will terminate. We describe the different cases now:
  \begin{itemize}
  \item[$\bullet$ Case 1:] The first case is if
    \begin{align}
    \nnf{F}({\mathbf{1}}^{k-2}, 1, 1, {\Psi}_{k+1}^n(\bY^*),{\mathbf{1}}^{k-2}, 0, 0, \neg {\Psi}_{k+1}^n(\bY^*),\bY^*)=0\\
    \text{and} \nnf{F}({\mathbf{1}}^{k-2}, 0, 1, {\Psi}_{k+1}^n(\bY^*),{\mathbf{1}}^{k-2}, 1, 0, \neg {\Psi}_{k+1}^n(\bY^*),\bY^*)=0
  \end{align}
  then, $x_{k-1}$ behaves as an AND gate, i.e.,
  \begin{align}
  \nnf{F}({\mathbf{1}}^{k-2}, x_{k-1}, 1, {\psi'}_{k+1}^n(\bY^*),{\mathbf{1}}^{k-2}, \bar{x}_{k-1}, 0, \neg {\psi'}_{k+1}^n(\bY^*),\bY^*)\leftrightarrow x_{k-1}\wedge \bar{x}_{k-1}
\end{align}
  which implies that $\zeta$ (with the Skolem functions assigned according to $\Psi$) will be satisfiable at $k-1$ and hence this terminates the proof.
  
\item[$\bullet$ Case 2:] This case is if
  \begin{align}
    \nnf{F}({\mathbf{1}}^{k-2}, 1, 1, {\Psi}_{k+1}^n(\bY^*),{\mathbf{1}}^{k-2}, 0, 0, \neg {\Psi}_{k+1}^n(\bY^*),\bY^*)=1
  \end{align}
  In this case, note that $\Psi_{k-1}(\bY*)=1$ and from Equation ~(\ref{eq:pf3}), we have
  \begin{align}
    \exists \bX_1^{k-2} \nnf{F}(\bX_1^{k-2},1,1,{\Psi}_{k+1}^n(\bY^*),\neg \bX_1^{k-2},0,0,\neg {\Psi}_{k+1}^n(\bY^*),\bY^*)=0
  \end{align}
  Thus the Skolem function $\Psi$ is wrong at level $k-1$, since it gives 1 but fixing $x_{k-1}=1$, there is no way to set lower variables to 1. In other words, we have reduced the problem by one level and can recursively apply this argument at level $k-1$.
\item[$\bullet$ Case 3:]
      \begin{align}
    \nnf{F}({\mathbf{1}}^{k-2}, 1, 1, {\Psi}_{k+1}^n(\bY^*),{\mathbf{1}}^{k-2}, 0, 0, \neg {\Psi}_{k+1}^n(\bY^*),\bY^*)=0=\Psi_{k-1}(\bY^*)\\
    \text{and } \nnf{F}({\mathbf{1}}^{k-2}, 0, 1, {\Psi}_{k+1}^n(\bY^*),{\mathbf{1}}^{k-2}, 1, 0, \neg {\Psi}_{k+1}^n(\bY^*),\bY^*)=1\\
    \text{i.e., }\nnf{F}({\mathbf{1}}^{k-2}, {\Psi}_{k-1}^n(\bY^*),{\mathbf{1}}^{k-2}, \neg {\Psi}_{k-1}^n(\bY^*),\bY^*)=1
      \end{align}
      Note that this case is possible only if $k-2\geq 1$. But if this is not the case, i.e., if $k-2=0$, and $\nnf{F}({\Psi}_{1}^n(\bY^*),{\mathbf{1}}^{k-2}, \neg {\Psi}_{1}^n(\bY^*),\bY^*)=1$, this implies that there exists no counter-example which contradicts Equation~(\ref{eq:pf0}). Now we have three subcases:
      \begin{itemize}
      \item[$\bullet$ Case 3(a):]
        \begin{align*}
          \nnf{F}({\mathbf{1}}^{k-3}, 1, {\Psi}_{k-1}^n(\bY^*),{\mathbf{1}}^{k-3}, 0, \neg {\Psi}_{k-1}^n(\bY^*),\bY^*)=0\\
          \nnf{F}({\mathbf{1}}^{k-3}, 0, {\Psi}_{k-1}^n(\bY^*),{\mathbf{1}}^{k-3}, 1, \neg {\Psi}_{k-1}^n(\bY^*),\bY^*)=0
        \end{align*}
        But this case reduces to Case 1 above, i.e., we can see that $x_{k-2}$ behaves as an AND gate (i.e., it is not $\wedge_{k-1}$-unrealizable), and so it terminates.
        \item[$\bullet$ Case 3(b):]
        \begin{align*}
          \nnf{F}({\mathbf{1}}^{k-3}, 1, {\Psi}_{k-1}^n(\bY^*),{\mathbf{1}}^{k-3}, 0, \neg {\Psi}_{k-1}^n(\bY^*),\bY^*)=1\\
          \text{and from~(\ref{eq:pf3}), we have, }\\
          \exists_1^{k-3}\nnf{F}(\bX_1^{k-3},1{\Psi}_{k-1}^n(\bY^*), \neg \bX_1^{k-3}, 0, \neg {\Psi}_{k-1}^n(\bY^*),\bY^*)=0
        \end{align*}
        which as in Case 2, reduces the problem by two levels.
      \item[$\bullet$ Case 3(c):]
        \begin{align*}
          \nnf{F}({\mathbf{1}}^{k-3}, 1, {\Psi}_{k-1}^n(\bY^*),{\mathbf{1}}^{k-3}, 0, \neg {\Psi}_{k-1}^n(\bY^*),\bY^*)=0\\
          \nnf{F}({\mathbf{1}}^{k-3}, 0, {\Psi}_{k-1}^n(\bY^*),{\mathbf{1}}^{k-3}, 1, \neg {\Psi}_{k-1}^n(\bY^*),\bY^*)=1
        \end{align*}
        But reduces to Case 3 at level $k-3$, thus ensuring strict progress in this case as well.        
        \end{itemize}
      Together this completes the proof.  
        \end{itemize}
\end{proof}

\subsection{Proofs from Section~\ref{sec:refine}}
\label{app:refine}

\skrefines*
\begin{proof}
Let $\bG(\bY)$ be a Skolem function vector for $\bX$ in
$\uF(\bX,\bY)$.  From condition (a) of Definition~\ref{def:refine}, we
know that $\forall\bY \left(\exists \bX F(\bX,\bY) \Rightarrow
{\uF}(\bG(\bY),\bY)\right)$.  Further, from condition (b) of
Definition~\ref{def:refine} and using $\bG(\bY)$ for $\bX'$, we have
$\forall\bY \left(\exists \bX F(\bX,\bY) \Rightarrow
F(\bG(\bY),\bY)\right)$.  This shows that $\bG(\bY)$ is a Skolem function
vector for $\bX$ in $F$.
\end{proof}

\refineprops*
\begin{proof}
\begin{enumerate}
\item The reflexivity of $\preceq_{syn}$ follows trivially from
Definition~\ref{def:refine}.  To see why $\preceq_{syn}$ is
transitive, suppose $F_1 \preceq_{syn} F_2$ and $F_2 \preceq_{syn}
F_3$. It follows from transitivity of $\Rightarrow$ that
$\forall \bY \left(\exists \bX F_3(\bX,\bY) \Rightarrow \exists \bX'
F_1(\bX,\bY)\right)$. This proves condition (a) of $F_1 \preceq_{syn} F_3$.
To prove condition (b) of $F_1 \preceq_{syn} F_3$, notice that
$\forall \bY \forall \bX'\left(\left(\exists \bX F_3(\bX,\bY) \wedge
F_1(\bX',\bY)\right) \Rightarrow \left(\exists \bX
F_3(\bX,\bY) \wedge \exists \bX'' F_2(\bX'',\bY) \wedge
F_1(\bX',\bY)\right)\right)$ by condition (a) of $F_2 \preceq_{syn}
F_3$.  Additionally, $\forall \bY \forall \bX'\left(\left(\exists \bX
F_3(\bX,\bY) \wedge \exists \bX'' F_2(\bX'',\bY) \wedge
F_1(\bX',\bY)\right)\Rightarrow \left(\exists \bX F_3(\bX,\bY) \wedge
F_2(\bX',\bY)\right)\right)$ by condition (b) of $F_1 \preceq_{syn}
F_2$.  Finally, by condition (b) of $F_2 \preceq_{syn} F_3$, it
follows that $\forall \bY \forall \bX'\left(\left(\exists \bX
F_3(\bX,\bY) \wedge F_2(\bX',\bY)\right) \Rightarrow
F_3(\bX',\bY)\right)$.  Putting all the parts together and by
transitivity of $\Rightarrow$, we have
$\forall \bY \forall \bX'\left(\left(\exists \bX F_3(\bX,\bY) \wedge
F_1(\bX',\bY)\right) \Rightarrow F_3(\bX',\bY)\right)$.  This proves condition
(b) of $F_1 \preceq_{syn} F_3$.

\item Suppose $\bigwedge_{y_j \in \bY}\left(F|_{y_j=0} \Leftrightarrow
  F|_{y_j=1}\right)$ and $\pi \models F(\bX,\bY)$. Then $F$ is
  semantically independent of $\bY$ and $\forall \bY
  F(\proj{\pi}{\bX}, \bY) = 1$ holds.  Therefore, $\forall \bY \exists
  \bX F(\bX,\bY) = 1$. Since $\forall \bY \exists \bX
  \form{\proj{\pi}{\bX}} = 1$ trivially, it follows that $\forall \bY
  \left(\exists \bX F(\bX,\bY) \Rightarrow \exists \bX'
  \form{\proj{pi}{\bX}}\right)$. Therefore condition (a) of
  Definition~\ref{def:refine} is satisfied.  Condition (b) of
  Definition~\ref{def:refine} follows from the observation that since
  $\pi \models F$ and $F$ is semantically independent of $\bY$, we
  have $\forall \bY \exists \bX F(\bX,\bY) = 1$ and $\forall \bY
  \forall \bX \form{\proj{\pi}{\bX}} \Rightarrow F(\bX,\bY)$.

\item Suppose $\bigwedge_{x_i \in \bX}\left(F|_{x_i=0} \Leftrightarrow
  F|_{x_i=1}\right)$.  Then $F$ is semantically independent of $\bX$.
  Substituting $1$ for $\uF$ in condition (a) of
  Definition~\ref{def:refine}, we get a tautology.  Similarly,
  substituting $1$ for $\uF$ in condition (b) of
  Definition~\ref{def:refine}, we get $\forall \bY \forall \bX'
  \left(\exists \bX F(\bX,\bY) \Rightarrow F(\bX',\bY)\right)$.
  Since $F$ is semantically independent of $\bX$, the above formula
  is also a tautology.  Hence condition (b) of Definition~\ref{def:refine}
  is also satisfied.

\item If $F$ is positive unate in $x_i$, then $F|_{x_i=0} \Rightarrow
  F|_{x_i=1}$.  It follows that $F \Leftrightarrow (\neg x_i \wedge
  F|_{x_i=0}) \vee (x \wedge F|_{x_i=1}) \Rightarrow F|_{x_i=1}$.
  Therefore, $\forall \bY \left(\exists \bX F(\bX,\bY) \Rightarrow
  \exists \bX' (x_i' \wedge F(\bX',\bY)|_{x_i'=1})\right)$.  This
  proves condition (a) of Definition~\ref{def:refine}.  To show that
  condition (b) of the definition also holds, note that $x_i' \wedge
  F(\bX',\bY)|_{x_i'=1} \Rightarrow F(\bX',\bY)$ is trivially a
  tautology.  The proof for the case when $F$ is negative unate in
  $x_i$ is analogous to the one above.

\item Suppose $\uF_1 \preceq_{syn} F_1$ and $\uF_2 \preceq_{syn} F_2$.
  \begin{enumerate}
  \item Since $\exists \bX \left(F_1(\bX,\bY) \vee F_2(\bX,\bY)\right)
    \Leftrightarrow \left(\exists \bX F_1(\bX,\bY) \vee \exists \bX
    F_2(\bX,\bY)\right)$ and $\exists \bX \left(\uF_1(\bX,\bY) \vee
    \uF_2(\bX,\bY)\right) \Leftrightarrow \left(\exists \bX
    \uF_1(\bX,\bY) \vee \exists \bX \uF_2(\bX,\bY)\right)$, condition
    (a) of Definition~\ref{def:refine} follows immediately.
    To see
    why condition (b) of the definition holds, notice that $\exists
    \bX \left(F_1(\bX,\bY) \vee F_2(\bX,\bY)\right) \wedge
    \left(\uF_1(\bX',\bY) \vee \uF_2(\bX',\bY)\right) \Rightarrow
    \left(\exists \bX F_1(\bX,\bY) \wedge \uF_1(\bX',\bY)\right) \vee
    \left(\exists \bX F_2(\bX,\bY) \wedge \uF_2(\bX',\bY)\right)$.  By
    condition (b) for $\uF_1\preceq_{syn} F_1$ and $\uF_2
    \preceq_{syn} F_2$, it follows that $\left(\exists \bX
    F_1(\bX,\bY) \wedge \uF_1(\bX',\bY)\right) \Rightarrow
    F_1(\bX',\bY)$ and $\left(\exists \bX F_2(\bX,\bY) \wedge
    \uF_2(\bX',\bY)\right) \Rightarrow F_2(\bX',\bY)$.  Hence, by
    transitivity of $\Rightarrow$, we get $\exists \bX
    \left(F_1(\bX,\bY) \vee F_2(\bX,\bY)\right) \wedge
    \left(\uF_1(\bX',\bY) \vee \uF_2(\bX',\bY)\right) \Rightarrow
    \left(F_1(\bX',\bY) \vee F_2(\bX',\bY)\right)$. Since this holds
    for all $\bY$ and $\bX'$, condition (b) of
    Definition~\ref{def:refine} is satisfied.

  \item Since the output supports of $F_1$ and $F_2$, and similarly of
    $\uF_1$ and $\uF_2$, are disjoint, we have $\exists \bX
    \left(F_1(\bX,\bY) \wedge F_2(\bX,\bY)\right) \Leftrightarrow
    \left(\exists \bX F_1(\bX,\bY) \wedge \exists \bX
    F_2(\bX,\bY)\right)$, and $\exists \bX \left(\uF_1(\bX,\bY) \wedge
    \uF_2(\bX,\bY)\right) \Leftrightarrow \left(\exists \bX
    \uF_1(\bX,\bY) \wedge \exists \bX \uF_2(\bX,\bY)\right)$.
    Therefore, condition (a) of Definition~\ref{def:refine} follows
    immediately.

    To see why condition (b) of the definition holds, notice that
    $\exists \bX \left(F_1(\bX,\bY) \wedge F_2(\bX,\bY)\right) \wedge
    \left(\uF_1(\bX',\bY) \wedge \uF_2(\bX',\bY)\right) \Rightarrow
    \left(\exists \bX F_1(\bX,\bY) \wedge \uF_1(\bX',\bY)\right) \wedge
    \left(\exists \bX F_2(\bX,\bY) \wedge \uF_2(\bX',\bY)\right)$.  By
    condition (b) for $\uF_1\preceq_{syn} F_1$ and $\uF_2
    \preceq_{syn} F_2$, it follows that $\left(\exists \bX
    F_1(\bX,\bY) \wedge \uF_1(\bX',\bY)\right) \Rightarrow
    F_1(\bX',\bY)$ and $\left(\exists \bX F_2(\bX,\bY) \wedge
    \uF_2(\bX',\bY)\right) \Rightarrow F_2(\bX',\bY)$.  Hence, by
    transitivity of $\Rightarrow$, we get $\exists \bX
    \left(F_1(\bX,\bY) \wedge F_2(\bX,\bY)\right) \wedge
    \left(\uF_1(\bX',\bY) \wedge \uF_2(\bX',\bY)\right) \Rightarrow
    \left(F_1(\bX',\bY) \wedge F_2(\bX',\bY)\right)$. Since this holds
    for all $\bY$ and $\bX'$, condition (b) of
    Definition~\ref{def:refine} is satisfied.
  \end{enumerate}
\end{enumerate}
\end{proof}

\refine*
\begin{proof}
  To prove part (\ref{lem:ref-further-1}), notice that $F \Rightarrow
  {\Def}_{\bT}$.  Hence, whenever $F(\bX,\bY)$ is satisfied, each of
  the functional definitions in ${\Def}_{\bT}$ are also satisfied.
  Therefore, condition (a) of Definition~\ref{def:refine} is
  satisfied.  For condition (b) of Definition~\ref{def:refine}, notice
  that for every value of $\bY$, only when the value of $\bX'$ is as
  given by ${\Def}_{\bT}(\bX',\bY)$, does ${\uF}(\bX',\bY)$ evaluate
  to $1$.  For these values of $\bX'$, if $\bY$ is such that
  $\exists \bX F(\bX,\bY)$ holds, then $F(\bX',\bY)$ must also hold
  since $\bX = \bT$ and $F \Rightarrow {\Def}_{\bT}$.

  To prove part (\ref{lem:ref-further-2}), consider
  $\theta_{F,\bT,x_i,0}$ to be a tautology; the proof for the case of
  $\theta_{F,\bT,x_i,1}$ being a tautology is analogous.  We show
  below that (a) $\forall \bY \left(\exists \bX
  F(\bX, \bY) \Rightarrow \exists \bX' (x_i' \wedge
  F(\bX', \bY)|_{x_i' =1})\right)$, and (b)
  $\forall \bY \forall \bX \left((x_i \wedge F(\bX,\bY)|_{x_i
  =1}) \Rightarrow F(\bX, \bY)\right)$.  Let $\sigma$ be an arbitrary
  element in $2^{|\bY|}$.  To see why (a) holds, suppose
  $F(\bX, \sigma) = 1$.  If $x_i = 1$, we set $\bX' = \bX$ and it
  follows that $(x_i \wedge F(\bX', \sigma)|_{x_i =1}) = 1$.  If $x_i
  = 0$, we set $x_j' = x_j$ for every $x_j \in \bX\setminus
  (\bT\cup\{x_i\})$, set $x_i'=1$ and set the value of every $x_j'$
  for $x_j \in \bT$ according its functional definition in
  ${\Def}_{\bT}(\bX',\bY)$.
  Since $\theta_{F,\bT,x_i,0}$ is a tautology, it follows
  that $(x_i' \wedge F(\bX', \sigma)|_{x_i' =1}) = 1$.  To see why
  (b) holds, suppose $(x_i \wedge F(\bX, \bY)|_{x_i =1}) = 1$. It
  follows trivially that $x_i$ must be set to $1$, and $F(\bX,\bY) = 1$.
\end{proof}

\refinesubsume*
\begin{proof}
 Observe that for any system
  of acyclic {\fdefs} $(\bT, {\Def}_{\bT})$ in $F$, since
  $F(\bX,\bY) \Rightarrow {\Def}_{\bT}$, the formula
  $\theta_{F,\bT,x_i,a}$ is a tautology iff
  $F(\bX,\bY)|_{x_i=a} \Rightarrow \exists \bT\,F(\bX,\bY)|_{x_i=1-a}$
  is a tautology.  It is now easy to see that if
  $\bT' \subseteq \bT \subseteq \bX$ and $\theta_{F,\bT',x_i,a}$ is
  valid, then $\theta_{F,\bT,x_i,a}$ is valid as well.
\end{proof}

\complete*
\begin{proof}[Proof of Theorem~\ref{thm:synnnf-complete}]
  The reverse direction is proved by first applying
  Theorem~\ref{thm:synnnf-exists-synth}(\ref{thm:synnnf-synth}) to
  $\uF$, and then noting that since $\uF \preceq_{syn} F$, every
  Skolem function vector for $\bX$ in $\uF$ is also a Skolem function
  vector for $\bX$ in $F$.  For the forward direction, let
  $\bpsi(\bY)$ be a Skolem function vector for $\bX$ in $F$ such that
  the size of an AND/OR/NOT gate circuit representation of $\bpsi$
  (denoted $|\bpsi|$) is polynomial in $|F|$.  As mentioned in
  Section~\ref{sec:prelim}, every such circuit can be converted to NNF
  in time $\bigO(|\bpsi|)$.  Hence the NNF representation of $\bpsi$
  is of size at most polynomial in $|F|$.  Therefore, w.l.o.g we
  consider $\bpsi$ to be in NNF.  Now consider the
  specification $\uF(\bX,\bY) \equiv \bigwedge_{i=1}^n \left((x_i
  \wedge \psi_i(\bY)) \vee (\neg x_i \vee \neg\psi_i(\bY))\right)$.
  Since no paths from $x_i$ and $\neg{x_i}$ ($x_i \in \bX$) meet at an
  $\wedge$-labeled node in the circuit representation of $\uF$, it
  follows that $\uF(\bX,\bY)$ is in {\syn}.  Furthermore, by
  construction of $\uF$, every Skolem function vector for $\bX$ in
  $\uF$ is necessarily component-wise semantically equivalent to
  $\bpsi$, which is itself a Skolem function vector for $\bX$ in $F$.
  Therefore, conditions (a) and (b) in Definition~\ref{def:refine} are
  satisfied by $\uF$, and hence $\uF \preceq_{syn} F$.
\end{proof}

\subsection{Proof from Section~\ref{sec:dsharp}}

\ctosyncorrect*
\begin{proof}
To see that {\CToSyn} always terminates, notice that every time the
recursion level $\ell$ in Algorithm~\ref{alg:c2syn} increases, the set
of output variables in the remaining set of clauses reduces by $1$.
Hence, the maximum value of $\ell$ can only be $|\bX|$, and the
recursion always terminates.  To see why {\FindFDOutsAndRefine}
(Algorithm~\ref{alg:fdoutsrefine}) terminates, notice that every time
$\bT'$ changes, its size increases by at least $1$, and hence $\bT'$
can change at most $|\bX|$ times.  Similarly, every time
$\mathcal{S}'$ changes, at least one variable is added to $\bT'$, and
hence $\mathcal{S}'$ cannot change more than $|\bX|$ times.

To see that the returned specification refines
$\cnfform{\mathcal{S}}$, notice that each of the \textbf{return}
statements in Algorithm~\ref{alg:c2syn} (i.e., lines $3$, $6$, $8$,
$12$, $17$ and $30$) uses one of the properties of refinement already
discussed in Section~\ref{sec:refine}.  Specifically, the correctness
of line $3$ is trivial.  The correctness of lines $6$ and $8$ use
Propositions~\ref{prop:refine-props}(\ref{prop:aa})
and \ref{prop:refine-props}(\ref{prop:aa}).  The correctness of lines
$12$ and $17$ use
Lemma~\ref{lem:refine-further}(\ref{lem:ref-further-1}).  The
correctness of line $30$ uses
Proposition~\ref{prop:refine-props}(\ref{prop:c}).
\end{proof}     

\end{document}